\documentclass[floatfix]{aastex61}
\pdfoutput=1 
\usepackage{amsmath,amstext}
\usepackage[T1]{fontenc}
\usepackage{apjfonts} 
\usepackage[figure,figure*]{hypcap}
\usepackage{enumitem}
\usepackage{multirow}

\shortauthors{D. Tak et al.}
\begin{document}
\title{Multiple Components in the Broadband $\gamma$-ray Emission of the Short GRB 160709A}
\author{Donggeun Tak}
\affiliation{Department of Physics, University of Maryland, College Park, MD20742, USA; \href{mailto:donggeun.tak@gmail.com}{donggeun.tak@gmail.com}}
\affiliation{NASA Goddard Space Flight Center, Greenbelt, MD 20771, USA}
\author{Sylvain Guiriec}
\affiliation{Department of Physics, The George Washington University, 725 21st Street NW, Washington, DC 20052, USA}
\affiliation{NASA Goddard Space Flight Center, Greenbelt, MD 20771, USA}
\author{Z. Lucas Uhm}
\affiliation{NASA Goddard Space Flight Center, Greenbelt, MD 20771, USA}
\author{Manal Yassine}
\affiliation{Istituto Nazionale di Fisica Nucleare, Sezione di Trieste, I-34127 Trieste, Italy}
\affiliation{Dipartimento di Fisica, Università di Trieste, I-34127 Trieste, Italy}
\author{Nicola Omodei}
\affiliation{W. W. Hansen Experimental Physics Laboratory, Kavli Institute for Particle Astrophysics and Cosmology, Department of Physics and SLAC National Accelerator Laboratory, Stanford University, Stanford, CA 94305, USA}
\author{Julie McEnery}
\affiliation{NASA Goddard Space Flight Center, Greenbelt, MD 20771, USA}
\affiliation{Department of Physics and Department of Astronomy, University of Maryland, College Park, MD 20742, USA}
\affiliation{Department of Physics, The George Washington University, 725 21st Street NW, Washington, DC 20052, USA}
\begin{abstract}
GRB 160709A is one of the few bright short gamma-ray bursts detected by both the Gamma-ray Burst Monitor and the Large Area Telescope on board the \textit{\textit{Fermi} Gamma-ray Space Telescope}. The $\gamma$-ray prompt emission of GRB 160709A is adequately fitted by combinations of three distinct components: (i) a nonthermal component described by a power law (PL) with a high-energy exponential cutoff, (ii) a thermal component modeled with a Planck function, and (iii) a second nonthermal component shaped by an additional PL crossing the whole $\gamma$-ray spectrum. While the thermal component dominates during $\sim$ 0.12 s of the main emission episode of GRB 160709A with an unusually high temperature of $\sim$ 340 keV, the nonthermal components dominate in the early and late time. The thermal component is consistent with the photospheric emission resulting in the following parameters: the size of the central engine, $R_{0}$ = $3.8 \substack{+5.9 \\ -1.8}$ $\times$ 10\textsuperscript{8} cm, the size of the photosphere, R$_{ph}$ = $7.4 \substack{+0.8 \\ -1.2}$ $\times$ 10\textsuperscript{10} cm, and a bulk Lorentz factor, $\Gamma$ = $728 \substack{+75 \\ -93}$ assuming a redshift of 1. The slope of the additional PL spectrum stays unchanged throughout the burst duration; however, its flux decreases continuously as a function of time. A standard external shock model has been tested for the additional PL component using the relation between the temporal and spectral indices (the closure relation). Each set of spectral and temporal indices from two energy bands (200 keV--40 MeV and 100 MeV--10 GeV) satisfies a distinct closure relation. From the closure relation test we derived the index for the electron spectral distribution, \textit{p} = 2.5 $\pm$ 0.1. The interaction of the jet with the interstellar environment is preferred over the interaction with the wind medium.
\end{abstract}

\section{Introduction}\label{sec:intro}
Gamma-ray bursts (GRBs) are the most energetic explosions in the universe. The GRBs are classified into two types based on their duration (T$_{\rm 90}$): long GRBs ($T_{\rm 90}$ > 2 s) from the collapse of massive stars \citep[e.g.,][]{Woosley1993, MacFadyen1999, Woosley2006} and short GRBs ($T_{\rm 90}$ $\lesssim$ 2 s) from the merger of two compact objects such as two neutron stars or a neutron star with a black hole \citep[e.g.,][]{Paczynski1986, Paczynski1991, Rosswog1999, GW170817}. Both long and short GRBs show similar spectral features. The observed keV--MeV prompt emission spectra of GRBs have been traditionally fitted with an empirical model, the Band function \citep{Band1993}, which is two smoothly connected power laws (PLs) with low- ($\alpha$) and high- ($\beta$) energy photon indices. Despite the empirical nature of the Band function, its low- and high-energy photon index values have been used to test the emission mechanisms as well as the underlying particle distributions. According to the standard internal shock model, accelerated electrons in relativistic collisionless shocks radiate via synchrotron radiation \citep{Rees1994}. The energy spectrum of the synchrotron radiation resulted from the PL electron distribution has an upper limit on its spectral steepness, $\alpha$ $\leq$ -2/3 \citep{Katz1994}. The observation of some hard spectra challenges any model invoking the synchrotron process as the main emission mechanism of the observed prompt emission spectrum \citep{Preece1998, Preece2000, Gruber2014}. For such hard spectra, modified internal shock models as well as alternative models have been introduced: a decaying magnetic field behind the shock front \citep{Peer2006, Derishev2007}, globally decreasing magnetic fields \citep{Uhm2014}, slow heating \citep{Asano2009}, synchrotron self-Compton scattering (SSC) \citep{Panaitescu2000}, and the Internal-collision-induced Magnetic Reconnection and Turbulence (ICMART) model \citep{Zhang2011}. In addition to exploring alternative emission mechansims, the observation of several GRBs suggests modification of the Band function itself. For several GRBs, the Band function was attenuated by a spectral cutoff at high energy in the sub-GeV band \citep[e.g.,][]{Catalog2013(LAT),Tang2015}. Furthermore, spectral cutoffs were reported at high energy in the $\lesssim$ MeV band \citep{Vianello2017} and at low energy in the X-ray energy band \citep{Oganesyan2017}.

Alternate spectral models to the Band function have been explored. For instance, a possible signature of thermal emission was explored to fit the data collected with the Burst And Transient Source Experiment (BATSE) on board the \textit{Compton $\gamma$-ray Observatory} (\textit{CGRO}) either by a single blackbody \citep{Ghirlanda2003} or a blackbody with an additional PL \citep{Ryde2004}. A sub-dominant thermal component in addition to a nonthermal one in the prompt emission were discovered from the long GRB 100724B \citep{Guiriec2011} and short GRB 120323A \citep{Guiriec2013} detected with the Gamma-ray Burst Monitor \citep[GBM;][]{GBM} on board the \textit{Fermi Gamma Ray Space Telescope} (hereafter, \textit{Fermi}). Similar results were reported in many short and long GRBs detected with \textit{Fermi} \citep{Axelsson2012, Guiriec2015a, Guiriec2015b}, \textit{CGRO}/BATSE \citep{Guiriec2016a}, \textit{Swift} and \textit{Suzaku} \citep{Guiriec2016b} and \textit{Wind}/Konus \citep{Guiriec2017}. These thermal components have been interpreted as emission from a photosphere. \cite{Meszaros2000} investigated the roles and possible observational features of the photosphere. The idea of the photospheric thermal emission has been extended with the consideration of various dissipative mechanisms \citep{Daigne2002,Rees2005,Beloborodov2010, Vurm2011}.

Using \textit{CGRO}/BATSE, \cite{Gonzalez2003} reported another deviation at high energy from the Band function adequately fitted by an additional PL. This result was confirmed using data collected with the Large Area Telescope \citep[LAT;][]{LAT} on board \textit{Fermi} for both long and short GRBs. The additional component generally crosses the whole gamma ray spectrum and extends up to a few GeV \citep[e.g.,][]{Abdo2009a, Ackermann2010, Ackermann2011, Yassine2017}. One possible explanation for this high-energy component is that it may originate from the external forward shock \citep[e.g.,][]{Ghirlanda2010, DePasquale2010}. The temporal and spectral properties of the external blast wave are well understood \citep{Sari1998,Zhang2006,Uhm2012}. According to the external forward shock model, the spectra and light curves of GRBs are characterized by a series of broken power laws \citep{Sari1998}, and the temporal and spectral indices are related by the "closure relation", which is determined by the surrounding environment \citep{Rees1994, Meszaros1997, Sari1998, Dai1998, Chevalier2000, Dai2001}. The LAT high-energy extended emission typically decreases as a function of time \citep{Ghisellini2010, Catalog2013(LAT)}, which supports the external forward shock model. Furthermore, the detailed spectral and temporal properties are consistent with those predicted by the external shock model \citep{Kumar2009, Kumar2010, DePasquale2010, Ghisellini2010, Panaitescu2017}, although bright and spiky temporal structures observed in the prompt phase of LAT light curves challenge this interpretation \citep{ZhangBB2011,Maxham2011}. Recently, \cite{Gompertz2018} applied the closure relation to a large sample of LAT-detected long GRBs and identified their surrounding environment conditions. Alternatively, the GeV emission has been interpreted as due to a magnetically dominated jet model based on the combined effects of dissipation from magnetic reconnection and nuclear collision \citep{Meszaros2011}. Synchrotron self-Compton \citep{Meszaros1993,Corsi2010} and hadronic emission \citep{Asano2009,Razzaque2010} have also been proposed.

Recently, \cite{Guiriec2015a} reported the simultaneous existence of the three components in GRBs detected with both \textit{Fermi}/GBM and \textit{Fermi}/LAT: (i) a thermal-like component interpreted as photospheric emission from the jet photosphere; (ii) a nonthermal component interpreted as the synchrotron emission from charged particles accelerated within the jet or as a strongly reprocessed photospheric emission; and (iii) a second nonthermal component whose physical process is unknown. \cite{Guiriec2016a} confirmed this result in a sample of GRBs detected with \textit{CGRO}/BATSE and \cite{Guiriec2016b} showed that this three-component model is a good fit to the whole broadband prompt emission spectrum of GRB 110205A from optical to $\gamma$-rays using \textit{Swift} and \textit{Suzaku} data.

In this paper, we present the analysis of the bright short GRB 160709A detected with both GBM and LAT. Even though the LAT has detected more than 120 GRBs, only 10\% of them are short GRBs \citep{Catalog2013(LAT)}. Due to the scarcity of short GRBs detected with LAT, the nature of their high-energy emission is not well established yet, and the bright short GRB 160709A will improve our understanding of the emission mechanisms powering short GRBs. The observation of GRB 160709A is presented in Section\ref{sec:observation}. Section~\ref{sec:analysis} reports GBM and LAT spectral analyses. The interpretations of the analysis results are discussed in Section~\ref{sec:discuss}. Finally, Section~\ref{sec:conclusion} summarizes and concludes the article.
\begin{figure*}[t]
 \centering
 \includegraphics[scale=0.6]{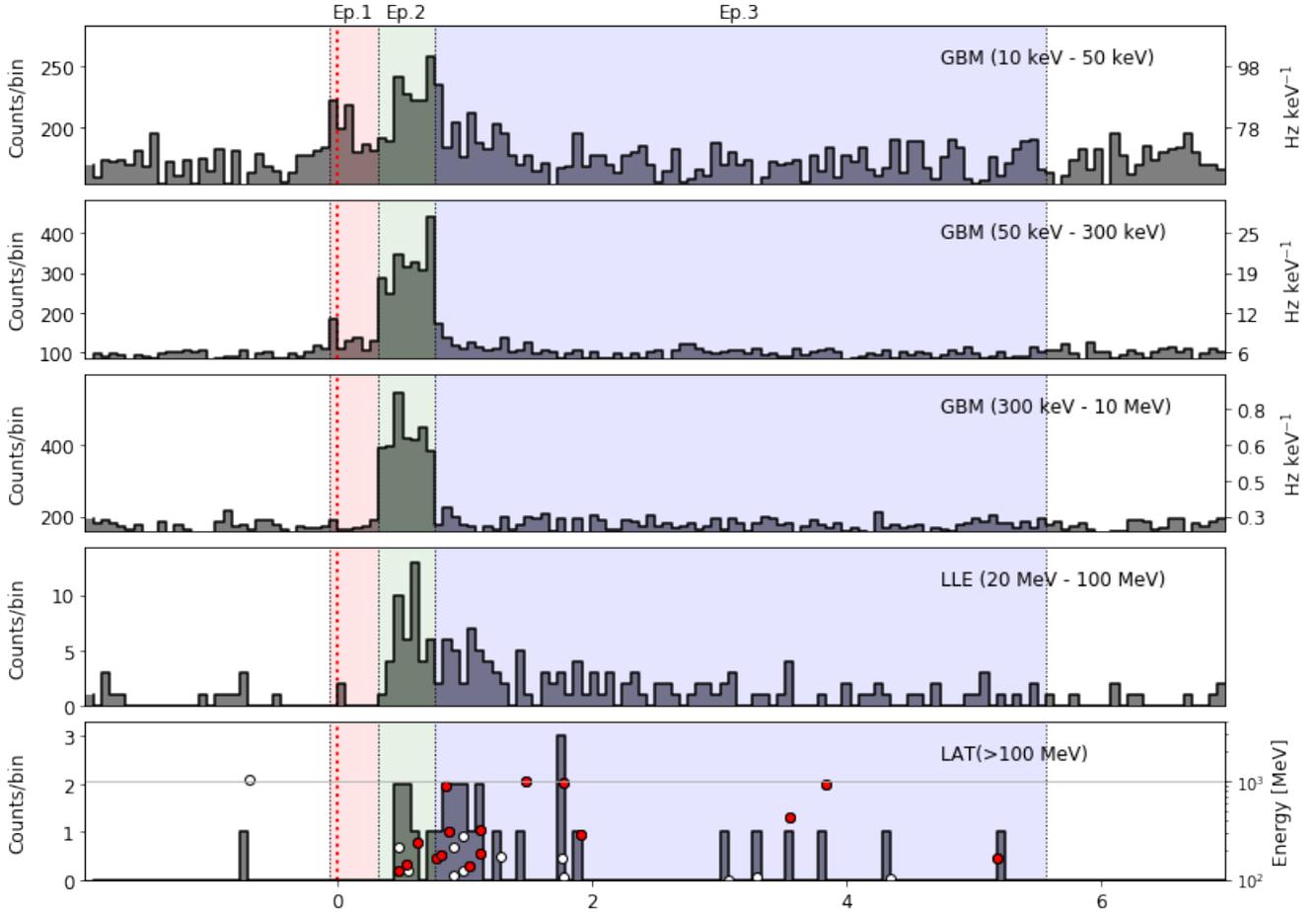}
 \caption{Composite light curve of GRB 160709A in different energy ranges with the GBM, LLE, and LAT data. The bin size is 0.064 seconds for all energy bands. The second panel shows the light curve in the energy range (50--300 keV) used for T\textsubscript{90} calculation. In the bottom panel, the energies of the LAT events are displayed as circles where the highly associated events with GRB 160709A (with probability $>$ 90\%) are filled in red. The red dotted vertical line represents GBM trigger time (T\textsubscript{0}). The time interval of T\textsubscript{90} is separated into three episodes by black dotted lines at T$_{0}$ + (-0.064, 0.320, 0.768, 5.568)s. The first and the last vertical lines represent T\textsubscript{05} and T\textsubscript{95}, respectively. The other two lines are chosen by the Bayesian-Block algorithm.}
 \label{fig:lc}
\end{figure*}
\section{Observation} \label{sec:observation}
At 19:49:03.50 UT on 2016 July 09 (hereafter $T$\textsubscript{0}), GRB 160709A triggered GBM with significance of 5.6 $\sigma$ \citep[GRB Coordinates Network (GCN) report,][]{2016GCN..19676...1J}. The ground analysis of \textit{Swift}/Burst Alert Telescope (BAT) reported a 9 $\sigma$ excess consistent with GRB 160709A at (R.A., decl.) = (235.996$^o$, -28.188$^o$) within a 0.04$^o$ error radius \citep[GCN report,][]{2016GCN..19681...1S}. GRB 160709A was also detected by LAT at $\sim$ 17 $\sigma$ with the clean event type of the data beyond 100 MeV \citep[GCN report,][]{2016GCN..19675...1G} as well as at 10 $\sigma$ using LAT Low Energy (hereafter, LLE) data beyond 10 MeV. The prompt emission of GRB 160709A was also detected by \textit{Wind}/Konus \citep[GCN report,][]{2016GCN..19677...1F}, the \textit{CALorimetric Electron Telescope (CALET)} \citep[GCN report,][]{2016GCN..19701...1A}, and \textit{AstroSat} \citep[GCN report,][]{2016GCN..19740...1B}. 

Figure~\ref{fig:lc} shows the light curves of GRB 160709A in various GBM and LAT energy bands. The GBM was triggered by a low-intensity pulse compared to the main emission episode observed about 0.3 s later. The main emission episode is followed by a $\sim$ 4s--long tail. The $T$\textsubscript{90} and $T$\textsubscript{50} durations of GRB 160709A computed between 50 and 300 keV \citep{Kouveliotou1993} using GBM data are 5.63 $\pm$ 1.29 s and 0.58 $\pm$ 0.20 s, respectively. This GRB is considered to belong to the short GRB category because its $T$\textsubscript{50} value is a characteristic of a short GRB (T\textsubscript{50} $\lesssim$ 1), although the T\textsubscript{90} value places GRB 160709A in the overlapped region of the $T_{90}$ distribution of long and short GRBs \citep{Catalog2016(GBM)}. This apparent inconsistency between $T$\textsubscript{90} and $T$\textsubscript{50} is explained by the fact that the former is more sensitive to the low energy tail that follows the main emission episode. A similar result was also reported for the very bright short GRB 090510 detected with GBM \citep{Ackermann2010}. The LAT detected more than 20 events above 100 MeV within $\sim$ 30 s. The highest energy photon is 991 MeV, detected at $T$\textsubscript{0} + 1.47 s. 
\begin{figure*}[t]
 \centering
 \includegraphics[scale=0.8]{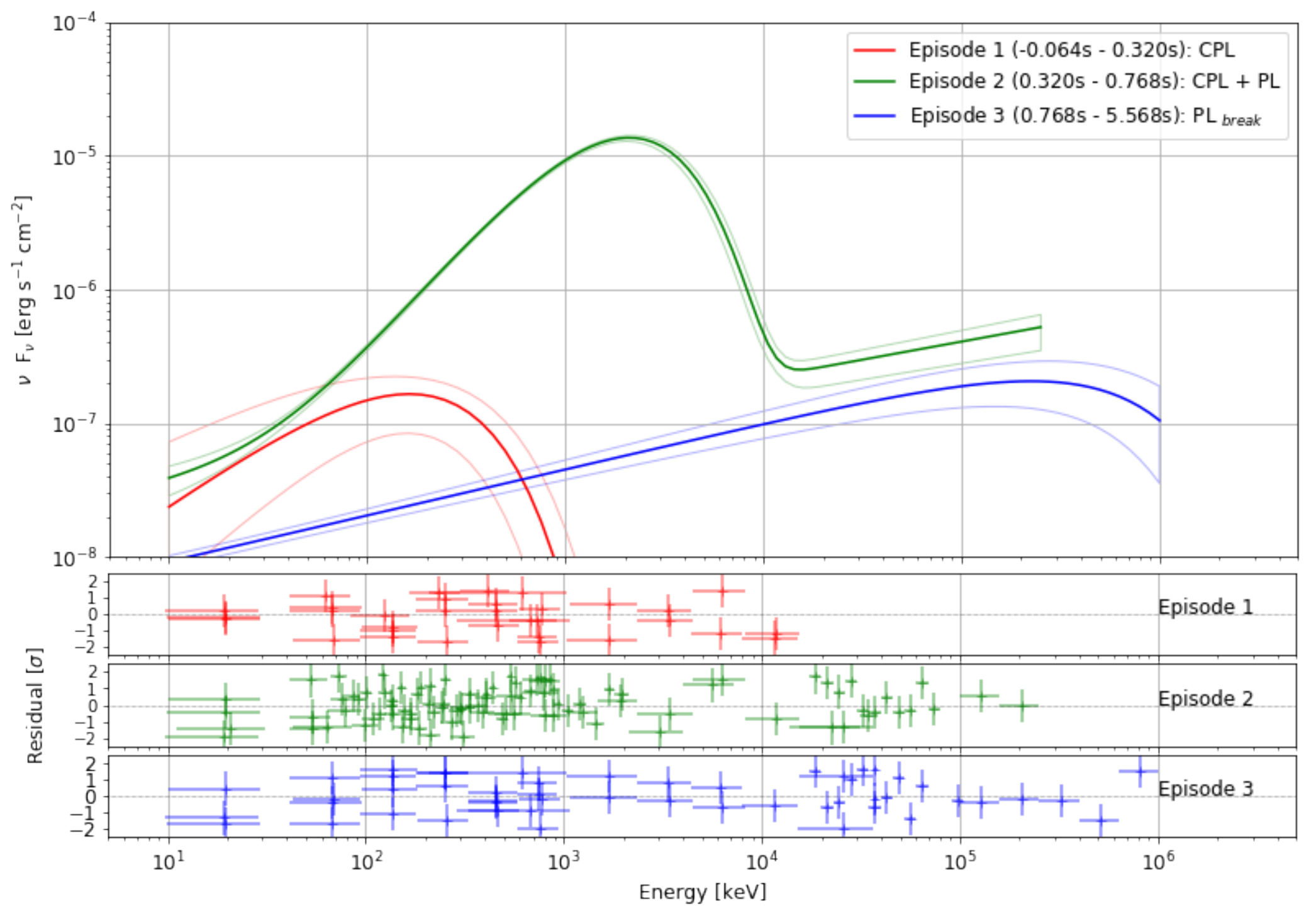}
 \caption{GRB 160709A spectral energy distributions as measured by the \textit{Fermi} GBM and LAT in different time intervals, using LAT Pass 8 data above 100 MeV and LAT--LLE data. Each solid curve represents the best-fit spectral shape, within the 1 $\sigma$ confidence level contour derived from the errors on the fit parameters. Each episode is painted with distinct colors in Figure~\ref{fig:lc}; red, green, and blue represent the first, second, and third episodes, respectively. The bottom panel shows the residuals of each best-fit model for the three episodes computed as (data-model)/error. Note that these residuals are not the same as the PG-stat contributions.}
 \label{fig:SED}
\end{figure*}

\section{Spectral Analysis} \label{sec:analysis}
\subsection{Analysis methods} \label{sec:methods}
We perform a spectral analysis of GRB 160709A using four GBM NaI detectors (n3, n4, n6, and n7, 8 keV--1 MeV), the two BGO detectors (b0 and b1, 200 keV--40 MeV) as well as LAT--LLE (20 MeV--100 MeV) and LAT--Transient020E (>100 MeV with a cut on the zenith angle at 100$^o$) data. We take the data of each NaI detector from the energy channels from 8 keV to the overflow channel (i.e., from channel 6 to channel 126) excluding channels close to the Iodine K-edge around 33.17 keV \citep[from channel 22 to channel 28;][]{Bissaldi2009}. Also, we select the energy channels for the two BGO detectors from 200 keV to the overflow channel (i.e., from channel 3 to channel 125). The background of each energy channel of each GBM detector was estimated by fitting background regions of the light curve before and after the burst with polynomial functions using \textit{RMfit 4.3.2}\footnote{\textit{RMfit}, \href{https://fermi.gsfc.nasa.gov/ssc/data/analysis/rmfit}{https://fermi.gsfc.nasa.gov/ssc/data/analysis/rmfit} for details.}.  A similar approach was used to estimate the background of LAT--LLE data.

To determine the GRB spectral shape, we test various models: simple PL, a PL with exponential cutoff (CPL), the Band function, a blackbody (BB) and the combinations of these models\footnote{\textit{Xspec} Model, \href{https://heasarc.gsfc.nasa.gov/xanadu/xspec/manual/Models.html}{https://heasarc.gsfc.nasa.gov/xanadu/xspec/manual/Models.html} for functional forms}. The spectral analysis is performed with \textit{Xspec 12.9.1}\footnote{\textit{Xspec}, \href{https://heasarc.gsfc.nasa.gov/xanadu/xspec}{https://heasarc.gsfc.nasa.gov/xanadu/xspec} for details.}. The fit parameters are estimated using the maximum likelihood. The best-fit model is determined by comparing the likelihood-based fitting statistics called PG-stat (Poisson data with a Gaussian background\footnote{PG-stat, \href{https://heasarc.gsfc.nasa.gov/xanadu/xspec/manual/XSappendixStatistics.html}{https://heasarc.gsfc.nasa.gov/xanadu/xspec/manual/XSappendix-Statistics.html} for details.}) and the degrees of freedom (dof) of each model. Basically, the model with the lowest PG-stat with the highest dof is the best-fit model. However, when a model has the lowest PG-stat with one degree of freedom lower than an alternative model, its PG-stat value must be less than the alternative one by at least 9 ($\Delta$PG-stat $>$ 9, which corresponds to $>$ 3 $\sigma$) to be chosen as a better description compared to the alternative model. As for the improvement of a $\Delta$PG-stat $<$ 9, we perform a simulation study to explore the best-fit model. For the simulation study, we use \textit{Xspec}--\textit{Fakeit} which is a tool for synthesizing bursts with a given model and its parameters including Poisson fluctuations. We generate 10\textsuperscript{5} synthetic spectra for each model and then search for the best model following the approach described in the Appendix of \cite{Guiriec2015a}.

We also perform an unbinned likelihood analysis of the LAT--Transient020E data alone in the extended time interval [$T$\textsubscript{0}, $T$\textsubscript{0} + 31 s] in the energy range from 100 MeV to 10 GeV with the \textit{Fermi Science Tool} version v11r5p3. We select the region of interest (ROI) within the radius of 15$^o$ from the \textit{Swift}/BAT localization. The backgrounds are estimated within 20$^o$ from the ROI center using the 3FGL sources\footnote{3rd Fermi $\gamma$-ray LAT point-source catalog, \href{https://fermi.gsfc.nasa.gov/ssc/data/access/lat/4yr\_catalog/}{https://fermi.gsfc.nasa.gov/ssc/data/access/lat/4yr\_catalog/}} \citep{3FGL2015} with `make3FGLxml.py'\footnote{make3FGLxml.py, \href{https://fermi.gsfc.nasa.gov/ssc/data/analysis/user/}{https://fermi.gsfc.nasa.gov/ssc/data/analysis/user/}}, the Galactic diffuse background with the recent `gll\_iem\_v06.fits'\footnote{gll\_iem\_v06.fits, \href{https://fermi.gsfc.nasa.gov/ssc/data/access/lat/BackgroundModels.html}{https://fermi.gsfc.nasa.gov/ssc/data/access/lat/BackgroundModels.html}} and the isotropic diffuse background with `iso\_P8R2\_TRANSIENT020E\_V6\_v06.txt' provided by the \textit{\textit{Fermi} Science Tool} package. The parameters of the 3FGL source backgrounds are fixed to the values given in the catalog \citep{3FGL2015} due to their weak contribution. 
\begin{table*}[t]
	\caption{Spectral fitting to GBM + LLE + LAT data (8 keV--10 GeV) for various time intervals}
	\centering 
	\begin{tabular}{c l c c c c c c c c c c}
  \hline\hline
 & & \multicolumn{3}{c}{Band Model} & & \multicolumn{2}{c}{Power Law with break} & & Blackbody & & \\\cline{3-5}\cline{7-8}\cline{10-10}
T - T\textsubscript{0}& Model &$\alpha$ & $\beta$ & $E_{\rm peak}$ & & $\Gamma$ & $E_{\rm break}$ & & kT &$PGstats$/dof&$\Delta PGstats$\\
$\left[\right.$s$\left.\right]$ & & & & $\left[\right.$keV$\left.\right]$& & &$\left[\right.$MeV$\left.\right]$ & & $\left[\right.$keV$\left.\right]$& \\\hline
\multicolumn{12}{c}{Time-integrated analysis for each episode}\\\hline
- 0.064 +0.320 	& PL 					& & & & & $-1.93\substack{+0.05 \\ -0.06}$ & & & & 710/709 &  \\
(Episode 1) & CPL		& $-0.95\substack{+0.38 \\ -0.31}$ & & $163\substack{+82 \\ -40}$ & & & & & & 683/708&- 29\\
\rule{0pt}{4ex}
 +0.320 +0.768 	& CPL		& $-0.40\substack{+0.05 \\ -0.05}$ & & $2329\substack{+149 \\ -139}$ & & & & & & 882/708 & \\
(Episode 2) & Band		& $-0.34\substack{+0.06 \\ -0.06}$ & $-3.17\substack{+0.11 \\ -0.12}$ & $2011\substack{+141 \\ -127}$ & & & & & & 753/707& - 129\\
				& CPL + PL		& $-0.19\substack{+0.08 \\ -0.07}$ & & $2097\substack{+120 \\ -114}$ & & $-1.73\substack{+0.04 \\ -0.04}$ & & & & 735/706 & - 147 \\
        & CPL + PL\textsubscript{break}		& $-0.18\substack{+0.08 \\ -0.08}$ & & $2081\substack{+120 \\ -114}$ & & $-1.61\substack{+0.07 \\ -0.06}$ & $58.6\substack{+46.0 \\ -20.0}$ & & & 726/705 & - 156 \\
			 	& CPL + BB		& $-1.46\substack{+0.06 \\ -0.06}$ & & $14690\substack{+3608 \\ -2619}$ & & & & & $351\substack{+17 \\ -16}$ & 832/706 &- 50 \\
        & 3CM\footnote{Three-component model (CPL + PL + BB)} & $-0.44\substack{+0.18 \\ -0.17}$ & & $2407\substack{+336 \\ -258}$ & & $-1.71\substack{+0.06 \\ -0.05}$ & & & $358\substack{+60 \\ -62}$ & 733/704 &- 149 \\
        & Fixed 3CM\footnote{Three-component model with fixed parameters suggested by \cite{Guiriec2015a}(with $\alpha$ of CPL = -0.7, $\Gamma$ of PL = -1.5, and BB)} & $-0.70$\textsubscript{fixed} & & $2785\substack{+233 \\ -234}$ & & $-1.50$\textsubscript{fixed} & & & $376\substack{+36 \\ -33}$ & 738/706 &- 144 \\
\rule{0pt}{4ex}
+0.768 +5.568	& PL 					& & & & & $-1.72\substack{+0.02 \\ -0.02}$ & & & & 857/709 & \\
(Episode 3)& PL\textsubscript{break}		& &	& & & $-1.66\substack{+0.03 \\ -0.03}$ & $219\substack{+134 \\ -67}$ & & & 845/708&- 12 \\
 & PL\textsubscript{break+low rolloff}\footnote{PL\textsubscript{break} with low energy exponential rolloff, E\textsubscript{rolloff} = $25\substack{+16 \\ -11}$ keV}		& &	& & & $-1.75\substack{+0.05 \\ -0.05}$ & $230\substack{+203 \\ -82}$ & & & 838/707&- 19 \\
        & PL\textsubscript{8 keV--200 keV}\footnote{GBM only}		& &	& & & $-1.22\substack{+0.16 \\ -0.15}$ & & & & 321/230& \\
        & PL\textsubscript{200 keV--100 MeV}\footnote{GBM + LLE}		& &	& & & $-1.77\substack{+0.08 \\ -0.07}$ & & & & 509/467& \\
        & PL\textsubscript{100 MeV--10 GeV}\footnote{LAT only}		& &	& & & $-2.22\substack{+0.24 \\ -0.24}$ & & & & & \\
				& PL\textsubscript{200 keV--10 GeV}\footnote{GBM + LLE + LAT}		& $-1.83\substack{+0.05 \\ -0.04}$ & & & & & & & & 521/477&\\
				& Broken PL\textsubscript{200 keV--10 GeV}		& $-1.77$\textsubscript{fixed} & $-2.18$\textsubscript{fixed} & & & & $166\substack{+207 \\ -95}$ & & & 517/477& - 4\\\hline\hline
\multicolumn{12}{c}{Time-resolved analysis for the second episode\footnote{Only selected time intervals are presented here. The other time intervals are fitted best with either a CPL or a CPL+PL model.}}\\\hline
+0.448 +0.512	& BB + PL 					& & & & & $-1.68\substack{+0.05 \\ -0.06}$ & & & $341\substack{+25 \\ -23}$ & 667/707 & \\
				& CPL + PL		& $0.91\substack{+0.38 \\ -0.32}$ & & $1423\substack{+148 \\ -129}$ & & $-1.67\substack{+0.06 \\ -0.06}$ & & & & 665/706 & -2 \\ 
+0.512 +0.576	& BB + PL 					& & & & & $-1.85\substack{+0.07 \\ -0.09}$ & & & $334\substack{+30 \\ -28}$ & 735/707 & \\
				& CPL + PL		& $0.72\substack{+0.42 \\ -0.35}$ & & $1422\substack{+200 \\ -156}$ & & $-1.85\substack{+0.08 \\ -0.09}$ & & & & 734/706 & -1 \\     
        +0.448 +0.576	& BB + PL 					& & & & & $-1.75\substack{+0.04 \\ -0.04}$ & & & $338\substack{+19 \\ -18}$ & 751/707 & \\
				& CPL + PL		& $0.85\substack{+0.28 \\ -0.24}$ & & $1419\substack{+115 \\ -101}$ & & $-1.74\substack{+0.04 \\ -0.05}$ & & & & 748/706 & -3 \\     
        & Fixed 3CM\footnote{$\Gamma$ in PL is fixed to $\Gamma$ = -1.73 based on the time-integrated result.} & $-0.70$\textsubscript{fixed} & & $3558\substack{+1883 \\ -1390}$ & & $-1.73$\textsubscript{fixed} & & & $338\substack{+23 \\ -21}$ & 746/706 &-5 \\
        \hline\hline
		\end{tabular}
 \label{table:fit}
\end{table*}

\subsection{Time-integrated spectral analysis for each episode} \label{sec:integrated}
We divide the time interval corresponding to $T$\textsubscript{90} into three episodes (Figure~\ref{fig:lc}). Both visual inspection of the composite light curve and the Bayesian-Block algorithm \citep{Scargle2013} in different energy bands reach the same conclusion and result in the following cuts. The first episode is the time interval [$T_{0}$ - 0.064 s, $T_{0}$ + 0.320 s] where the emission is noticeable only in the low-energy region. The second time interval [$T_{0}$ + 0.320 s, $T_{0}$ + 0.768 s] corresponds to the main episode and consists of the bulk of the prompt emission in both GBM and LLE energy ranges. The third episode [$T_{0}$ + 0.768 s, $T_{0}$ + 5.536 s] corresponds to the decaying phase of the keV--MeV emission tail with high-energy photons detected above 100 MeV. The burst peak is detected almost simultaneously in the GBM (10 keV--5 MeV) and LLE (20 MeV--100 MeV) energy ranges; however, it is delayed by $\sim$ 0.3 s in the LAT range ($>$ 100MeV). We perform spectral analyses in each episode. Table~\ref{table:fit} reports the fit results of various models listed in the previous section (see Appendix~\ref{app:fit}, for the agreement between data and the best-fit spectral model).
\subsubsection{The first episode} \label{sec:first}
The best-fit model for the first episode [$T$\textsubscript{0} - 0.064 s, $T$\textsubscript{0} + 0.320 s] is CPL. This model show a significant improvement in the PG-stat compared to the PL with $\Delta$PG-stat $\sim$ 29 which corresponds to 5.2 $\sigma$. The index of CPL is $-0.95\substack{+0.38 \\ -0.31}$, in agreement with the typical low-energy index of GRBs, with $E_{\rm peak}$ = $163\substack{+82 \\ -40}$ keV. Even though the best-fit model is CPL, it is possible that we underestimate its complexity due to the low fluence of this episode.
\subsubsection{The second episode (main peak)} \label{sec:second}
We find some interesting features in this time interval [$T$\textsubscript{0} + 0.320 s, $T$\textsubscript{0} + 0.768 s]. The Band function, the standard GRB model, is not sufficient for describing the prompt emission spectral shape of GRB 160709A. Instead, the CPL + PL model ($\alpha = -0.19\substack{+0.08 \\ -0.07}$; $E_{\rm peak} = 2097\substack{+120 \\ -114}$ keV; $\Gamma = -1.73\substack{+0.04 \\ -0.04}$), is a better description of data than the Band function alone ($\Delta$PG-stat $\sim$ 18, $\sim$ 4.2 $\sigma$). We try to fit the data with the Band + PL model, but then $\beta$ becomes very soft ($\leq$ -10) so that the Band function and CPL are indistinguishable.

We add a break in the PL component of the CPL + PL model (CPL + PL\textsubscript{break}). The spectral fit yields a break energy at $E_{\rm break}$ = $59 \substack{+46 \\ -20}$ MeV with a decrease of the PG-stat value by 9. The improvement of PG-stat, however, is not enough to confirm the existence of the break in the PL component.

To check the existence of a blackbody component (BB), we test CPL + BB to the data. The value of PG-stat obtained with this model increases by $\sim$ 50 with respect to CPL alone, therefore, the CPL + BB model is less preferred than the CPL + PL model with the same number of dof. We also test a model consisting of three components; CPL + PL + BB. This model fits the data similarly as the CPL + PL model with only $\Delta$PG-stat $\sim$ 2 for two dof. As suggested by \cite{Guiriec2015a}, we fit the data with the three-component model fixing the two spectral parameters, $\alpha$ of CPL and $\Gamma$ of PL to -0.7 and -1.5, respectively. Both this model and the CPL + PL model give similar PG-stat values ($\Delta$ PG-stat $\sim$ 3) with the same dof, although the three-component model returns much bigger errors with respect to the CPL + PL model.

We perform simulation studies with \textit{Xspec}--\textit{Fakeit} for all alternative models (CPL + PL, CPL + PL\textsubscript{break}, CPL + PL + BB and CPL + PL + BB with fixed parameters). As a result, we find that all models resemble each other under the observed brightness of GRB 160709A, and one can reproduce others' results within 1 $\sigma$. Under the circumstances, we conclude that the CPL + PL is enough to explain the observed data and choose the model as the best-fit model for this episode. 

\subsubsection{The third episode (weak tail of the prompt emission)} \label{sec:third}
A weak tail of the prompt emission is observed during the third episode. In this time interval, the PL\textsubscript{break} model is the best-fit model (see Figure~\ref{fig:SED} and Figure~\ref{fig:fit1}). The fit result shows a break at 219$\substack{+134 \\ -67}$ MeV with $\Delta$PG-stat = 12 ($\sim$ 3.5 $\sigma$) with respect to a simple PL model. On the other hand, we reanalyze this episode with the GBM and LLE data (GBM + LLE, 8 keV--10 GeV) and find the hint of a break at $E_{\rm break}$ = $314\substack{+285 \\ -115}$ MeV ($\Delta$PG-stat = 7). This confirms that the break does not originate from the systematic effects such as different effective areas between LLE and LAT. We also test a model, PL\textsubscript{break} with an exponential rolloff at low energy\footnote{see \href{https://heasarc.gsfc.nasa.gov/xanadu/xspec/manual/node234.html}{https://heasarc.gsfc.nasa.gov/xanadu/xspec/manual/node234.html}. Note that the cutoff and rolloff models have the same functional form but they are referred to by different names in order to distinguish cutoffs in high- and low-energy spectra.}. This model yields the lowest PG-stat, but the $\Delta$PG-stat compared to PL\textsubscript{break} is not enough to compensate for one dof ($\Delta$PG-stat = 7). Neither the Band function + low energy rolloff nor a broken PL + low-energy rolloff show any improvement compared to PL\textsubscript{break} + low-energy rolloff.

We also perform energy-resolved spectral analyses for this episode with the PL model in the energy bands; 8 keV--200 keV, 200 keV--100 MeV and 100 MeV--10 GeV (Table~\ref{table:fit}). For 100 MeV--10 GeV analysis, we use the \textit{\textit{Fermi} Science Tool} rather than \textit{Xspec} as the former provides a better background estimation for LAT data. The three energy bands are well-fitted with PL, with clearly different photon indices, -1.22 (8 keV--200 keV), -1.77 (200 keV--100 MeV) and -2.22 (100 MeV--10 GeV) (Table~\ref{table:fit}). We double-check the LAT result with \textit{Xspec} which shows similar results within 1 $\sigma$. The data in the large energy range (200 keV--10 GeV) is fitted also with a broken PL model. The low and high spectral indices of this model are fixed to the values obtained from the PL fits in (200 keV--100 MeV) and (100 MeV--10 GeV) ranges, -1.77 and -2.22, respectively. The break energy of the broken PL is consistent with $E_{\rm break}$ of PL\textsubscript{break}, which is the best-fit model of the third time interval in the energy band from 8 keV to 10 GeV.

\subsection{Detailed temporal and spectral analyses} \label{sec:resolved}
\subsubsection{Time-resolved spectral analysis of the main emission} \label{sec:resolvedSecond}
\begin{figure}[t]
\centering
 \includegraphics[scale=0.5]{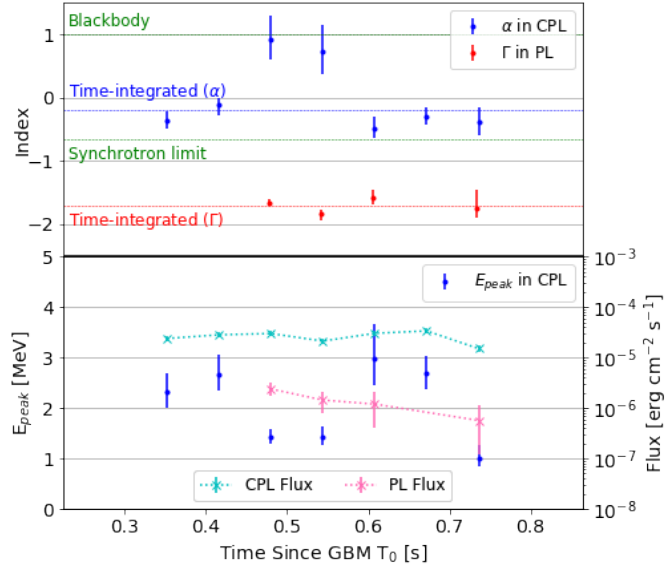}
 \caption{Evolution of the CPL and PL parameters during the main emission episode. The top panel shows the evolutions of CPL index ($\alpha$, blue) and PL index ($\Gamma$, red). The horizontal lines show the time-integrated best-fit values of $\alpha$ (blue) and $\Gamma$ (red), and blackbody radiation and synchrotron radiation limit (green, 1 and -2/3, respectively). The bottom panel shows the evolution of $E$\textsubscript{peak} of CPL (Blue) and the fluxes of CPL and PL in 10 keV--40 MeV, which are cyan and pink respectively.}
 \label{fig:evol_idxEpeak}
\end{figure}
\begin{figure}[t]
 \centering
 \includegraphics[scale=0.6]{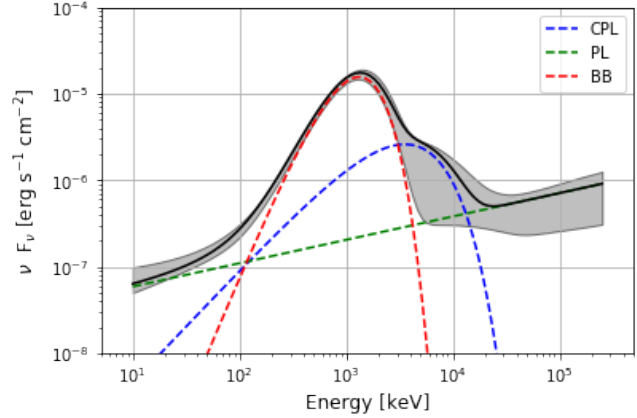}
 \caption{Spectral energy distributions of GRB 160709A in the thermal dominant phase [T\textsubscript{0} + 0.448 s, T\textsubscript{0} + 0.576 s] for the joint GBM + LLE + LAT analysis. The solid black curve represents the CPL + PL + BB within 1 $\sigma$ confidence level contour. The temperature of the BB component is about 340 keV.}
 \label{fig:bbcomp}
\end{figure}
We perform a time-resolved analysis in the main episode with 64 ms time-binned intervals with the GBM + LLE + LAT data. Since the CPL + PL model is the best-fit model in the time-integrated interval, we test the model in each time interval. When it is difficult to constrain PL due to a low fluence in the high-energy regime, we use CPL alone.

Figure~\ref{fig:evol_idxEpeak} shows the time evolution of PL and CPL parameters. The PL index ($\Gamma$) is almost constant at $\sim$ -1.7 for all time intervals (red in Figure~\ref{fig:evol_idxEpeak}). This value is consistent with the time-integrated result, $\Gamma$ $\sim$ -1.73 (Table~\ref{table:fit}). The PL flux reaches its maximum during the third time interval and then decreases with time (pink in Figure~\ref{fig:evol_idxEpeak}).

The CPL parameters (blue and cyan in Figure~\ref{fig:evol_idxEpeak}) show significant evolution with time. Particularly, in the third and fourth time intervals, [$T_{0}$ + 0.448 s, $T_{0}$ + 0.512 s] and [$T_{0}$ + 0.512 s, $T_{0}$ + 0.576 s], the slope of CPL ($\alpha$) reaches $\sim$ +0.9 and $\sim$ +0.8, respectively. Such values are similar to a typical BB spectral index. We test the BB + PL model in these two time intervals. The model successfully fits the data with $\Delta$PG-stat $\lesssim$ 2 with one more dof with respect to the CPL + PL model (Table~\ref{table:fit}). The temperatures of BB in the two sequential time intervals are $341\substack{+25 \\ -23}$ keV and $334\substack{+30 \\ -28}$ keV, respectively. The evolution of the temperature seems to be natural. Considering the evolution of $\alpha$ and the fit result with the BB + PL model in the these sequential time intervals, there is a thermal dominant phase lasting $\sim$ 0.12 s, [$T$\textsubscript{0} + 0.448 s, $T$\textsubscript{0} + 0.576 s]. In other time intervals, $\alpha$ is almost constant around -0.3, which is harder than the expected index of the synchrotron emission process as reported in many observations \citep[e.g.,][]{Katz1994, Preece1998, Preece2000, Gruber2014}.

The parameter $E_{\rm peak}$ of CPL fluctuates during the main episode (bottom panel in Figure~\ref{fig:evol_idxEpeak}). Particularly, in the time intervals where $\alpha$ is very hard ($\sim$ 0.9), the values of $E_{\rm peak}$ tend to be lower than those of the other time intervals. This implies that the $E_{\rm peak}$ fluctuation may be related to the BB component existing in the thermal dominant phase, [$T_{0}$ + 0.448 s, $T_{0}$ + 0.576 s]. The spectral analysis in the thermal dominant phase is performed with the three-component model (CPL + PL + BB) as shown in Figure~\ref{fig:bbcomp}. The peak energy of CPL and the temperature of BB are $E_{\rm peak}$ $\sim$ 3600 keV and kT $\sim$ 340 keV, respectively (Table~\ref{table:fit}). If BB with kT = 340 keV is fitted with CPL, the temperature of BB is equivalent to $E_{\rm peak}$ $\sim$ 1300 keV, $E_{\rm peak}$ $\sim$ 3.9 kT. This $E_{\rm peak}$ of the BB component is consistent with the $E_{\rm peak}$ of CPL ($\sim$ 1400 keV) in the the third and fourth time intervals, [T$_{0}$ + 0.448 s, T$_{0}$ + 0.512 s] and [T$_{0}$ + 0.512 s, T$_{0}$ + 0.576 s]. The fluxes of the CPL and BB components in the energy range from 10 keV to 10 MeV are (0.6 $\pm$ 0.4) $\times$ 10$^{-5}$ erg cm$^{-2}$ s$^{-1}$ and (2.1 $\pm$ 0.3) $\times$ 10$^{-5}$ erg cm$^{-2}$ s$^{-1}$, respectively. Depending on the relative dominance of the two components, $E_{\rm peak}$ of CPL in the CPL + PL model is constrained to one of the two peaks; i.e., as the thermal (nonthermal) component dominates the nonthermal (thermal) component, $E_{\rm peak}$ is constrained to the peak energy of thermal (nonthermal) component. The $E_{\rm peak}$ fluctuation is, therefore, self-consistent with the existence of the BB component in the second episode. 

\subsubsection{Spectral and temporal analyses of the extended emission}
\begin{figure}[t]
\centering
 \includegraphics[scale=0.5]{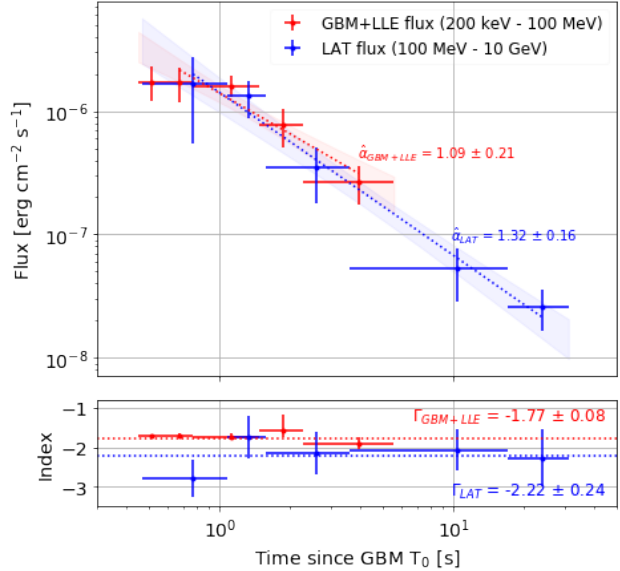}
 \caption{Temporal evolution of the flux and the spectral index during the extended emission in GBM + LLE (Red, 200 keV--100 MeV) and LAT (Blue, 100 MeV--10 GeV) energy ranges. The top panel shows the flux temporal evolution in dashed lines within the 1 $\sigma$ confidence level contour derived from the errors on the fit parameters. The indices of the temporal decay are displayed in this panel. The bottom panel shows the temporal evolution of PL spectral indices.}
 \label{fig:plcomp}
\end{figure}
In the third episode, a spectral break at $\sim$ 200 MeV is observed (Section~\ref{sec:third}). To track the temporal evolution of the energy spectrum below and above $E_{\rm break}$ independently, we perform spectral and temporal analyses in two different energy bands, GBM + LLE (200 keV to 100 MeV) and LAT (100 MeV to 10 GeV), with \textit{Xspec} and \textit{\textit{Fermi} Science Tool}, respectively. We ignore energy bands below 200 keV due to a steep break in the lower-energy band at $\sim$ 150 keV; i.e., NaI low-energy channels, 1 - 71, are ignored. For the GBM + LLE analysis, the time interval [$T_{0}$ + 0.448 s, $T_{0}$ + 5.568 s], is considered. This time interval contains the part of the second episode and the third episode, assuming that the additional PL component observed in the second episode and the PL\textsubscript{break} component in the third episode originate from the same source. As for the LAT analysis, we use the time interval, [$T_{0}$ + 0.464 s, $T_{0}$ + 31 s] where they are the time of the first and last LAT events associated with GRB 160709A (with probability $>$90\%). 

Figure~\ref{fig:plcomp} (top panel) shows the evolution of the PL fluxes in the two energy bands. Both GBM + LLE and LAT show time-decaying features except for the first point in the GBM + LLE energy range. We perform the maximum likelihood fit with the following PL equation:
\begin{equation}
\begin{aligned}
F(t) = F_{0} \left(\frac{t}{1s}\right)^{- \hat{\alpha}}.
\end{aligned}
\end{equation}
The flux in the two energy ranges decreases as a function of time, with marginally different exponents for the temporal decays: the temporal indices $\hat{\alpha}$ of the GBM + LLE and LAT analyses are 1.09 $\pm$0.21 and 1.32 $\pm$ 0.16, respectively. During the time-decaying phase, the photon indices in the GBM + LLE and LAT energy bands are also different: 1.77 $\pm$ 0.08 and 2.22 $\pm$ 0.24, respectively (bottom panel in Figure~\ref{fig:plcomp}). 

The photon index of GBM + LLE in the third episode is similar to that of the additional PL component observed in the second episode. Also, the flux of GBM + LLE decreases with time (Figure~\ref{fig:plcomp}), similar to the evolution of the flux of the additional PL component (Figure~\ref{fig:evol_idxEpeak}). These analogous features suggest that an emission process producing a PL spectrum is continuous from the main episode to the tail end of the emission (at least in the energy band from 200 keV to 100 MeV). 

\section{Interpretation and Discussion} \label{sec:discuss}
\begin{figure}[t]
 \includegraphics[scale=0.6]{BBresult.pdf}
 \centering
 \caption{Estimation of physical parameters as a function of redshift; the initial fireball radius R\textsubscript{0} $[f_{N-th}/\epsilon_{T}]^{-3/2}$, the photospheric radius $R_{\rm ph}$ $[(1+\sigma_{m})f_{N-th}]^{1/4}$ and the Lorentz factor $\Gamma$ $[(1+\sigma_{m})f_{N-th}]^{1/4}$. They were calculated by the photospheric model \citep{Daigne2002, Peer2007, Hascoet2013}. The solid-black line represents the most probable value at a given redshift within 1 $\sigma$ confidence level contour. The three parameters increase with the redshift. For redshift z = 1, the size of the central engine is $R_{0}$ = $3.8 \substack{+5.9 \\ -1.8}$ $\times$ 10\textsuperscript{8} cm, the size of the photosphere is R$_{ph}$ = $7.4 \substack{+0.8 \\ -1.2}$ $\times$ 10\textsuperscript{10} cm, and the bulk Lorentz factor is $\Gamma$ = $728 \substack{+75 \\ -93}$ assuming that the ratio $f_{N-th}/\epsilon_{T}$ and the product $(1+\sigma_{m})f_{N-th}$ are equal to unity each.}
 \label{fig:photosphere}
\end{figure}
\subsection{Thermal emission component}
Given the steepness of the spectrum in the time-resolved analysis of the main episode, we interpret the spectrum as being dominated by thermal emission during $\sim$ 0.12 s. Photospheric emission is naturally expected from the standard fireball model \citep[e.g.,][]{Piran1999}, and the photospheric model is suggested to explain the very hard low-energy spectral index (thermal emission) \citep{Meszaros2000, Meszaros2002, Rees2005, Peer2006b, Asano2013}. Since we observe a very hard spectral index in GRB 160709A in the time interval [$T_{0}$ + 0.448 s, $T_{0}$ + 0.576 s], we test the photospheric model. This model constrains several important physical parameters such as the size of the central engine $R$\textsubscript{0}, the photospheric radius $R$\textsubscript{ph} and the Lorentz factor $\Gamma$ \citep{Daigne2002, Peer2007, Hascoet2013}. They are expressed as
\begin{equation} \label{eqR0}
R_{0} \simeq \left[\frac{D_{L}\mathcal{R}}{2(1+z)^{2}} \left( \frac{F_{Th}}{F_{N-th}} \right) ^{3/2}\right]\times\left[\frac{f_{N-th}}{\epsilon_{T}}\right]^{3/2},
\end{equation}
\begin{equation} \label{eqRph}
R_{ph} \simeq \left[\frac{\sigma_{T}}{16m_{p}c^{3}}\frac{D_{L}^{5}\mathcal{R}^{3}F_{N-th}}{(1+z)^{6}} \right]^{1/4}\times[(1+\sigma_{m})f_{N-th}]^{-1/4},
\end{equation}
and
\begin{equation} \label{Lorentz}
\Gamma \simeq \left[\frac{\sigma_{T}}{m_{p}c^{3}}\frac{(1+z)^{2}D_{L}F_{N-th}}{\mathcal{R}} \right]^{1/4}\times[(1+\sigma_{m})f_{N-th}]^{-1/4},
\end{equation}
where $z$ and $D$\textsubscript{L} are redshift and luminosity distance. The luminosity distance $D$\textsubscript{L} is calculated as
\begin{equation} \label{DL}
D_{L} = (1+z)\frac{c}{H_{0}}\int_{0}^{z}\frac{dz'}{\sqrt{\Omega_{m}(1+z')^{3}+\Omega_{\Lambda}}}
\end{equation}
with cosmological parameters from \cite{Planck2016}. $F$\textsubscript{Th} and $F$\textsubscript{N-th} are the measured fluxes from the thermal component (BB) and nonthermal components (CPL and PL), respectively. The fluxes of the thermal and nonthermal components are (2.1 $\pm$ 0.3) $\times$ 10$^{-5}$ erg cm$^{-2}$ s$^{-1}$ and (0.9 $\pm$ 0.5) $\times$ 10$^{-5}$ erg cm$^{-2}$ s$^{-1}$, respectively, which are computed in the energy band from 8 keV to 250 MeV. $\epsilon_{T}$ is the fraction of the initial energy released by the source in thermal form, and $\sigma_{m}$ is the magnetization of the relativistic outflow at the end of the acceleration process. $f$\textsubscript{Nth} is the efficiency of the nonthermal emission mechanism observed in the spectrum. $\mathcal{R}$ represents the ratio between the measured BB flux and the expected BB flux by Stefan-Boltzmann law; 
\begin{equation}
\mathcal{R}= \left(\frac{F_{BB}}{\sigma T_{BB}^4}\right)^{1/2}.
\end{equation}

As the redshift of GRB 160709A is unknown, therefore, we estimate $R$\textsubscript{0}, $R$\textsubscript{ph} and Lorentz factor ($\Gamma$) as a function of redshift (Figure~\ref{fig:photosphere}). The parameter $R$\textsubscript{0} ranges from 10\textsuperscript{7}--10\textsuperscript{9} cm, $R$\textsubscript{ph} from 10\textsuperscript{8}--10\textsuperscript{11} cm, and $\Gamma$ from 10\textsuperscript{2} to a few 10\textsuperscript{3}, assuming that the ratio $f_{Nth}/\epsilon_{T}$ and the product $(1+\sigma_{m})f_{Nth}$ are of order unity, each, for 0.01 $\lesssim$ z $\lesssim$ 10. This assumption is possible when $f_{Nth}$ and $\epsilon_{T}$ have the same order. The $f_{N-th}$ term should be $f_{Nth}$ $\lesssim$ 0.1 for an internal shock model \citep{Daigne1998} and can be higher than this limit for other models such as the magnetic reconnection model \citep[see e.g.,][]{Spruit2001, Lyutikov2003, Zhang2011}. In terms of the fraction of the thermal energy of the source, the standard GRB scenario suggests $\epsilon_{T}$ $\lesssim$ 0.1 \citep{Daigne2002, Zhang2009,Guiriec2011,Hascoet2013}. The magnetization factor $\sigma_{m}$ in equations ~\ref{eqRph} and ~\ref{Lorentz} varies widely from $\sigma_{m}$ $\lesssim$ 1 to $\sigma_{m}$ $\gg$ 1 depending on the theoretical models; however, its impact on $R$\textsubscript{ph} and $\Gamma$ are negligible. A variation of $\sigma_{m}$ of order 10\textsuperscript{4} results in negligible changes of $R$\textsubscript{ph} and $\Gamma$, of order 10. \\

Since GRB 160709A is a short GRB, the expected size of $R_{0}$ is 10--100 times larger than the Schwarzschild radius of the central engine ($\sim$ 10\textsuperscript{6} cm) \citep{Ryde2009,Peer2015}. On the other hand, the size of photosphere $R_{\rm ph}$ is known to be 10\textsuperscript{11}--10\textsuperscript{13} cm. The typical redshift of short GRBs is z $\sim$ 1 \citep{Racusin2011}, and with a redshift of this order (or above) the values of $R_{0}$ and $R_{\rm ph}$ become plausible, $R_{0}$ $\sim$ 3.8 $\times$ 10$^{8}$ cm and $R_{\rm ph}$ $\sim$ 7.4 $\times$ 10$^{10}$ cm. We note that they are still small, but can be explained by the unexpectedly high temperature of the BB component (kT $\sim$ 340 keV). The Lorentz factor for z $\gtrsim$ 1 extends from a few 10\textsuperscript{2} to a few 10\textsuperscript{3} in harmony with a bulk Lorentz factor constrained from $\gamma\gamma$ opacity in other GRBs \citep{Lithwick2001,Fenimore1993, Baring1997, Abdo2009b,Abdo2009c, Racusin2011, Zhao2011}. Testing the photosphere model for the observed thermal emission, we suggest that the redshift of GRB 160709A is z $\gtrsim$ 1 to be consistent with the theoretical expectations and other observations. 

\subsection{Additional PL component\\ and high-energy extended emission}
\begin{figure}[t]
 \includegraphics[scale=0.7]{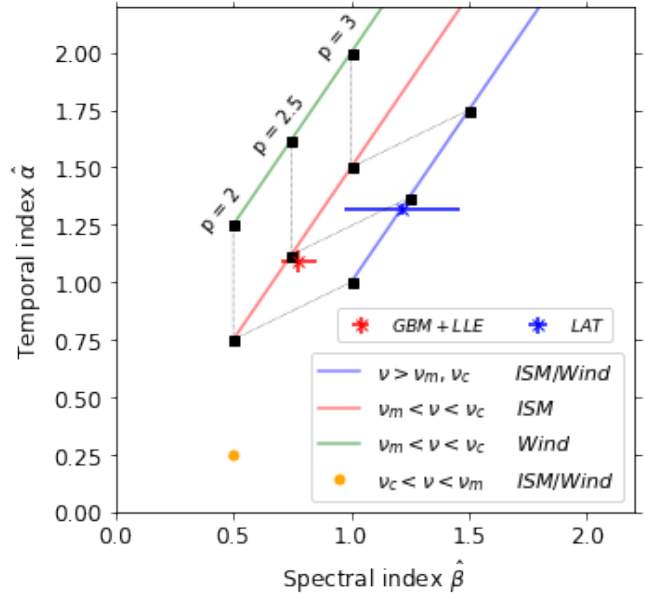}
 \centering
 \caption{Closure relation test. The red and blue crosses correspond to the set indices obtained with the GBM + LLE and LAT analyses, respectively. The three colored lines (blue, red, and green) correspond to various closure relations. The gray lines connect the black points that have the same electron energy index (\textit{p} = 2, 2.5, and 3) of the various closure relations. A weighted average of the electron spectral index of the two data sets is \textit{p} = 2.49 $\pm$ 0.12. The best-matched closure relation for the GBM + LLE analysis is given by $\nu_{m} < \nu < \nu_{c}$ in the ISM environment (solid red line), while the LAT analysis supports the relation given by $\nu > \nu_{c}, \nu_{m}$ (solid blue line).}
 \label{fig:closure_relation}
\end{figure}

\begin{table*}[t]
	\caption{Closure relation test}
	\centering 
	\begin{tabular}{c c c c c c c c c}
  \hline\hline
    & Energy Range & $\hat{\alpha}$ & $\hat{\beta}$ & $\textit{p}_{\hat{\alpha}}$ & $\textit{p}_{\hat{\beta}}$ & Weighted Average $\textit{p}$\\\hline
GBM + LLE	& 200 keV--100 MeV  & $1.09 \pm 0.21$ & $0.77\pm0.08$ & $2.46 \pm 0.29$ & $2.55 \pm 0.16$ & \multirow{2}{*}{$2.49 \pm 0.12$}\\
LAT	& 100 MeV--10 GeV  & $1.32 \pm 0.16$ & $1.22\pm0.24$  & $2.43 \pm 0.21$ & $ 2.44 \pm 0.49$ \\
		\hline\hline
	\end{tabular}
  	\label{table:closure_relation}
\end{table*}

The additional spectral component emerges 0.448 s after $T_{0}$ during the main emission episode and persists up to the GeV domain until the end of the extended emission (Section~\ref{sec:resolvedSecond}). The flux in two energy bands, GBM + LLE (200 keV--100 MeV) and LAT (100 MeV--10GeV), decreases with time as a PL function with two different temporal exponents (see Figure~\ref{fig:plcomp}). This PL decaying feature is well explained by the standard external forward shock model \citep[e.g.,][]{Kumar2009, Kumar2010, DePasquale2010, Ghisellini2010, Panaitescu2017}. \cite{Sari1998} calculated the expected broadband spectrum of synchrotron emission and its corresponding light curve using a PL distribution of electrons in an expanding relativistic shock. The evolution of the spectrum and light curve of a GRB depends on its surrounding environment, its jet geometry, the electron distribution in the jet, and the cooling condition of the accelerated electrons \citep{Rees1994, Meszaros1997, Sari1998, Dai1998, Chevalier2000, Dai2001, Gao2013}. The two representative surrounding medium profiles for GRBs, which are described as a function of radius, are the interstellar medium (ISM, $n(r) =$ constant) and the wind ($n(r) \propto$ r$^{-2}$) profiles. The synchrotron cooling regime is determined by the two physical parameters $\nu_{m}$ and $\nu_{c}$ \citep{Sari1998}. The parameter $\nu_{m}$ corresponds to the minimum Lorentz factor of the electrons accelerated in the shocks, and $\nu_{c}$ is the frequency of electrons after cooling down in shocks. If $\nu_{m}$ is larger than $\nu_{c}$, all accelerated electrons cool down rapidly (fast cooling). The opposite case is called the slow cooling case, and the accelerated electrons with $\nu$ > $\nu_{c}$ can cool down. For electrons with a PL distribution of index \textit{p} and a spectral flux $F_{\nu} = F_{\nu, 0}$ $t^{- \hat{\alpha}}$ $\nu^{- \hat{\beta}}$, the spectral index $\hat{\beta}$ where $\hat{\beta}$ = -($\Gamma$+1), and the temporal index $\hat{\alpha}$ can be expressed as a function of \textit{p} depending on the surrounding environment and cooling conditions. In the ISM environment, the index $\alpha$ is either -1/4 ($\nu_{c}$ < $\nu$ < $\nu_{m}$) or (2 - 3 $\times$ \textit{p})/4 ($\nu$ > $\nu_{m}$, $\nu_{c}$) for the fast cooling case, and either 3 $\times$ (\textit{p}-1)/4 ($\nu_{m}$ < $\nu$ < $\nu_{c}$) or (2 - 3 $\times$ \textit{p})/4 ($\nu$ > $\nu_{m}$, $\nu_{c}$) for the slow cooling case. Likewise, the index $\hat{\beta}$ is either 1/2 ($\nu_{c}$ < $\nu$ < $\nu_{m}$) or \textit{p}/2 ($\nu$ > $\nu_{m}$, $\nu_{c}$) for the fast cooling case, and either (\textit{p}-1)/2 ($\nu_{m}$ < $\nu$ < $\nu_{c}$) or \textit{p}/2 ($\nu$ > $\nu_{m}$, $\nu_{c}$) for the slow cooling case. The dependence of $\hat{\beta}$ and $\hat{\alpha}$ with the electron spectral index \textit{p} therefore shows that the two parameters are related by a closure relation.

We test the forward shock model with a set of closure relations derived in an adiabatic hydrodynamic evolution \citep[for the complete set of closure relations, see][]{Gao2013}. Figure~\ref{fig:closure_relation} shows the closure relations of different surrounding environment and cooling regime conditions. The two sets of spectral and temporal indices from the GBM + LLE and LAT analyses seem to be related with different closure relations. The set obtained from GBM + LLE analysis is positioned at a closure relation of the slow cooling condition ($\nu_{m}$ < $\nu$ < $\nu_{c}$) and ISM environment (red line in Figure~\ref{fig:closure_relation}). On the other hand, the set from the LAT analysis is at a closure relation corresponding to the fast cooling condition ($\nu$ $>$ $\nu_{m}$ and $\nu_{c}$) with an undetermined surrounding environment condition (blue line in Figure~\ref{fig:closure_relation}). 

From the two $\hat{\beta}$ and $\hat{\alpha}$ sets and corresponding closure relations, we compute the value of the electron spectral index \textit{p} of GRB 160709A (Table~\ref{table:closure_relation}). The \textit{p} values from the two different closure relations are consistent with each other. A weighed average of $p_{\beta}$ is 2.49 $\pm$ 0.12, which is consistent with other observational studies \citep[e.g.,][]{Shen2006, Starling2008, Curran2009, Curran2010}.

The closure relation from the GBM + LLE implies that the surrounding medium of GRB 160709A is the ISM environment rather than the wind environment. This accords with the theoretical expectation for the surrounding medium of short GRBs \citep{Paczynski1986,Paczynski1991,Rosswog1999,Abbott2017}. 

The closure relation of the GBM + LLE energy band implies $\nu_{\rm GBM + LLE}$ < $\nu_{c}$, while that of the LAT energy band suggests $\nu_{\rm LAT}$ > $\nu_{c}$. The two results indicate that $\nu_{c}$ is located in between the GBM + LLE energy band and the LAT energy band. The spectral break detected with 3.5 $\sigma$ in the third episode ($E_{\rm break}$ $\sim$ 166 MeV) supports the hypothesis of the existence of $\nu_{c}$ between the two energy bands. We explore the plausibility of the high $\nu_{c}$ with a simple assumption. Assuming that the GRB isotropic energy $E$\textsubscript{iso} = 10\textsuperscript{52} erg and surrounding medium density $n$ = 10\textsuperscript{-3} cm\textsuperscript{-3} \citep{Bernardini2007, Caito2009}, the electron cooling frequency $\nu_{c}$ in the ISM environment can be expressed as \citep{Sari1998}
\begin{equation}
\begin{aligned}
&\nu_{c} \simeq 7.9 \times 10^{14} \epsilon_{B}^{-3/2} \left(\frac{E\textsubscript{iso}}{10^{52}ergs}\right)^{-1/2} \left(\frac{n}{10^{-3}cm^{-3}}\right)^{-1} \left(\frac{t}{1s}\right)^{-1/2}\ Hz\\.
\end{aligned}
\end{equation}
If $\nu_{c}$ is 166 MeV at 3 s, the middle of the third episode, the above equation gives the fraction of the total energy contained in the magnetic field, $\epsilon_{B} \simeq 5.1 \times 10^{-4}$. This fraction is consistent with other observations \citep[e.g.,][]{Santana2014,Kumar2009,Kumar2010}. \cite{Santana2014} made a distribution of $\epsilon_{B}$ with \textit{Swift} observations and found that $\epsilon_{B}$ has a range from 10\textsuperscript{-8} to 10\textsuperscript{-3}.

\section{Summary and Conclusion} \label{sec:conclusion}
GRB 160709A is an unusual short burst detected by both GBM and LAT, which allows us to analyze the burst in broad energy ranges. GRB 160709A triggered GBM with a weak soft emission. This emission is followed by the main emission lasting $\sim$ 1 s. A weak tail of the prompt emission continues about 3 s after the main episode, which increases GBM $T$\textsubscript{90}. The GRB spectrum consists of several components including the thermal component; CPL, PL, and BB. The multi-component spectrum implies that GRB 160709A is produced not by a single emission process but with multiple emission processes including both thermal and nonthermal emissions. In summary, our observational analyses of the prompt emission of GRB 160709A show the following features:
\begin{itemize}[labelwidth=3.5cm]
 \item[1.]We determined that the keV--GeV spectrum of GRB 160709A is well described by a combination of CPL and PL. The weak soft emission triggering GBM is described by CPL with $\alpha$ $\sim$ -1 and $E_{\rm peak}$ $\sim$ 163 keV. For the main emission, the best-fit model is the CPL + PL model, although many other models well-fitted the data including the three-component model (CPL + BB + PL). After the main emission episode, a simple PL is observed continuously up to the end of the LAT extended emission.
 \item[2.]In time-resolved analyses of the main emission episode, we found very steep spectra in the sequential time intervals and regarded them as a thermal dominated phase lasting $\sim$ 0.12 s. In this phase, a CPL component was observed with very hard spectral index $\sim$ +0.9, therefore, it can be completely replaced by a BB component. The temperature of BB is about 340 keV, which is unexpectedly high compared to other sub-dominant thermal emission reports. The time-resolved analysis suggests that the best description of GRB 160709A spectrum in the main episode is the combination of the three components (CPL + PL + BB).
 \item[3.]The observation of high-energy events (LAT) is delayed compared to the lower-energy events (GBM). The delayed emission seems to be related to the onset of the additional PL component, which starts to appear during the main emission and remains until the end of the prompt phase. The best-fit model of this time interval is PL\textsubscript{break} with 3.5 $\sigma$ significance. The flux of this component decreases as a function of time (PL), and the spectral index of this component does not significantly change in time. The two energy bands (GBM + LLE and LAT) show different spectral and temporal indices.
 \item[4.]\cite{Guiriec2010} reported a detailed analysis of three of the brightest short GRBs detected with \textit{Fermi}/GBM. The sub-dominant thermal component observed in GRB 160709A was reported for the first time in 2011 in a long GRB \citep{Guiriec2011} and in 2013 in a short GRB \citep{Guiriec2013}, thus after the publication of \cite{Guiriec2010}. It is not excluded that such a photospheric emission is also present in the three short GRBs of \cite{Guiriec2010} as suggested in a new analysis not yet published. Therefore, GRB 160709A may not be that different from the bursts published in \cite{Guiriec2010}. \cite{Guiriec2013} reported the existence of an intense thermal component ($kT$ $\sim$ 10 -- 13 keV) in one of the brightest short GRBs detected with \textit{Fermi}/GBM; in this article and conversely to GRB 160709A, there was no clear indication of the existence of an extra PL in addition to the other two components.
  
\end{itemize}
The interpretations of these observational features lead to several conclusions.
\begin{itemize}[labelwidth=3.5cm]
 \item[1.]A thermal component in the time-resolved analyses can be interpreted as the photosphere emission of GRB 160709A. This burst clearly shows the thermal emission (BB) with very high temperature overwhelming nonthermal emission (CPL) in the 100 - 1000 keV energy band during the time interval [$T_{0}$ + 0.448 s, $T_{0}$ + 0.576 s]. Assuming the redshift of GRB\,160709A $\gtrsim$ 1, we derived the central engine radius $R$\textsubscript{0} $\simeq$ 10\textsuperscript{8} cm and the photospheric radius 10\textsuperscript{11} cm, which is surprisingly small compared to other previous reports.
 \item[2.] The origin of the extended emission is regarded as the external forward shock. The spectral and temporal indices of the GBM + LLE (200 keV--100 MeV) and the LAT (100 MeV--10 GeV) analyses separately have different closure relations, although the implications from the two closure relation do not conflict. We inferred that $\nu_{c}$ seems to be located between GBM + LLE and LAT energy bands. The surrounding environment is preferred to be the ISM environment rather than the wind environment. This analysis revealed the electron spectral index \textit{p} = 2.49 $\pm$ 0.12 for GRB 160709A. This is the first trial of testing GRB closure relations in the two distinct high-energy bands, and the observational properties are well interpreted by the external forward shock model. 
\end{itemize}

\acknowledgments
\section*{Acknowledgement}
The \textit{Fermi} LAT Collaboration acknowledges generous ongoing support from a number of agencies and institutes that have supported both the development and the operation of the LAT as well as scientific data analysis. These include the National Aeronautics and Space Administration and the Department of Energy in the United States, the Commissariat \`a l'Energie Atomique and the Centre National de la Recherche Scientifique / Institut National de Physique Nucl\'eaire et de Physique des Particules in France, the Agenzia Spaziale Italiana and the Istituto Nazionale di Fisica Nucleare in Italy, the Ministry of Education, Culture, Sports, Science and Technology (MEXT), High Energy Accelerator Research Organization (KEK) and Japan Aerospace Exploration Agency (JAXA) in Japan, and the K.~A.~Wallenberg Foundation, the Swedish Research Council and the Swedish National Space Board in Sweden.
 
Additional support for science analysis during the operations phase is gratefully acknowledged from the Istituto Nazionale di Astrofisica in Italy and the Centre National d'\'Etudes Spatiales in France. This work performed in part under DOE
Contract DE-AC02-76SF00515.
\software{RMfit (4.3.2), XSPEC (v12.9.1; \cite{Arnaud1996}), Fermi Science Tools (v11r5p3)}

\begin{figure}[h]
 \includegraphics[scale=0.7]{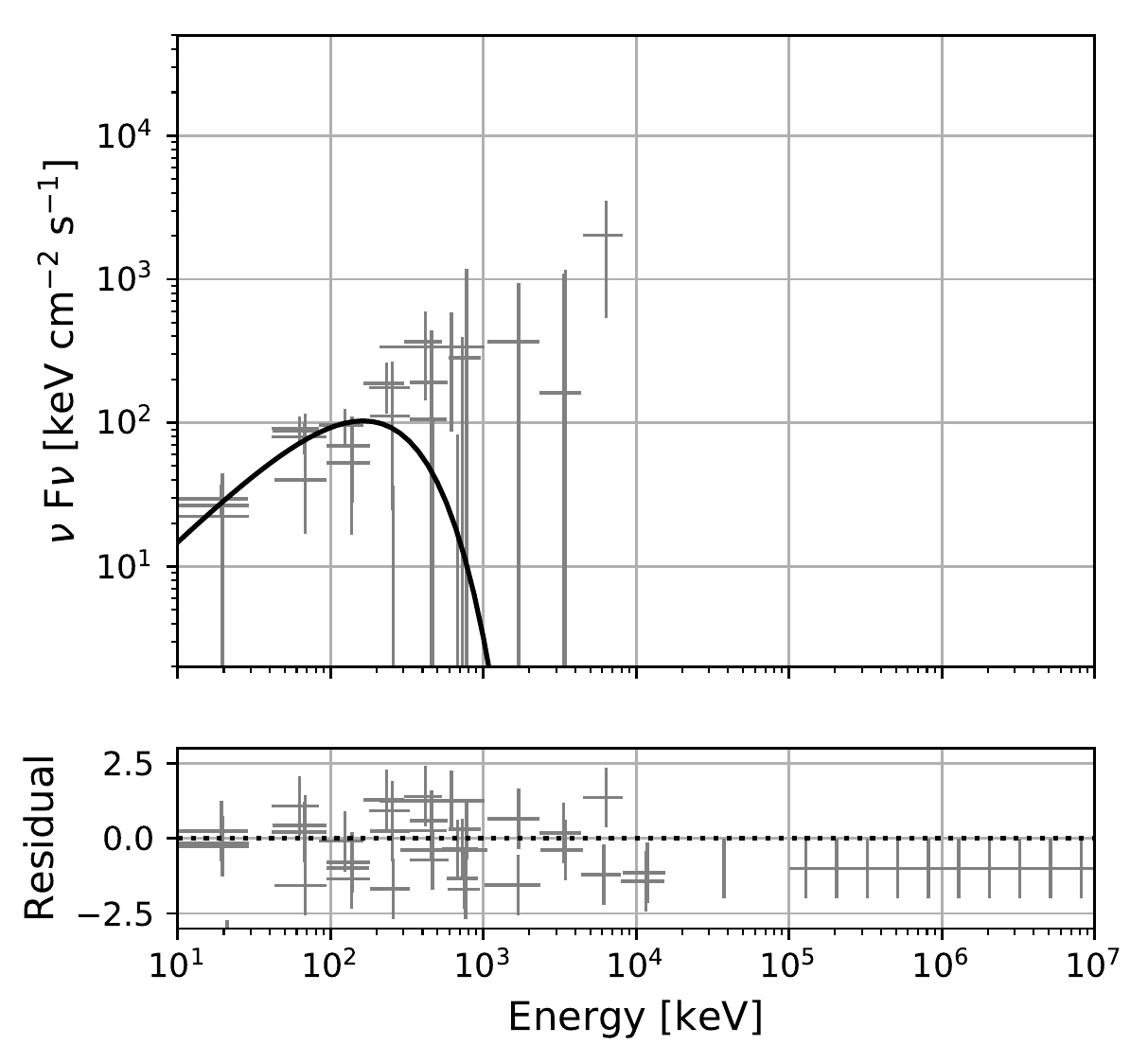}
 \includegraphics[scale=0.7]{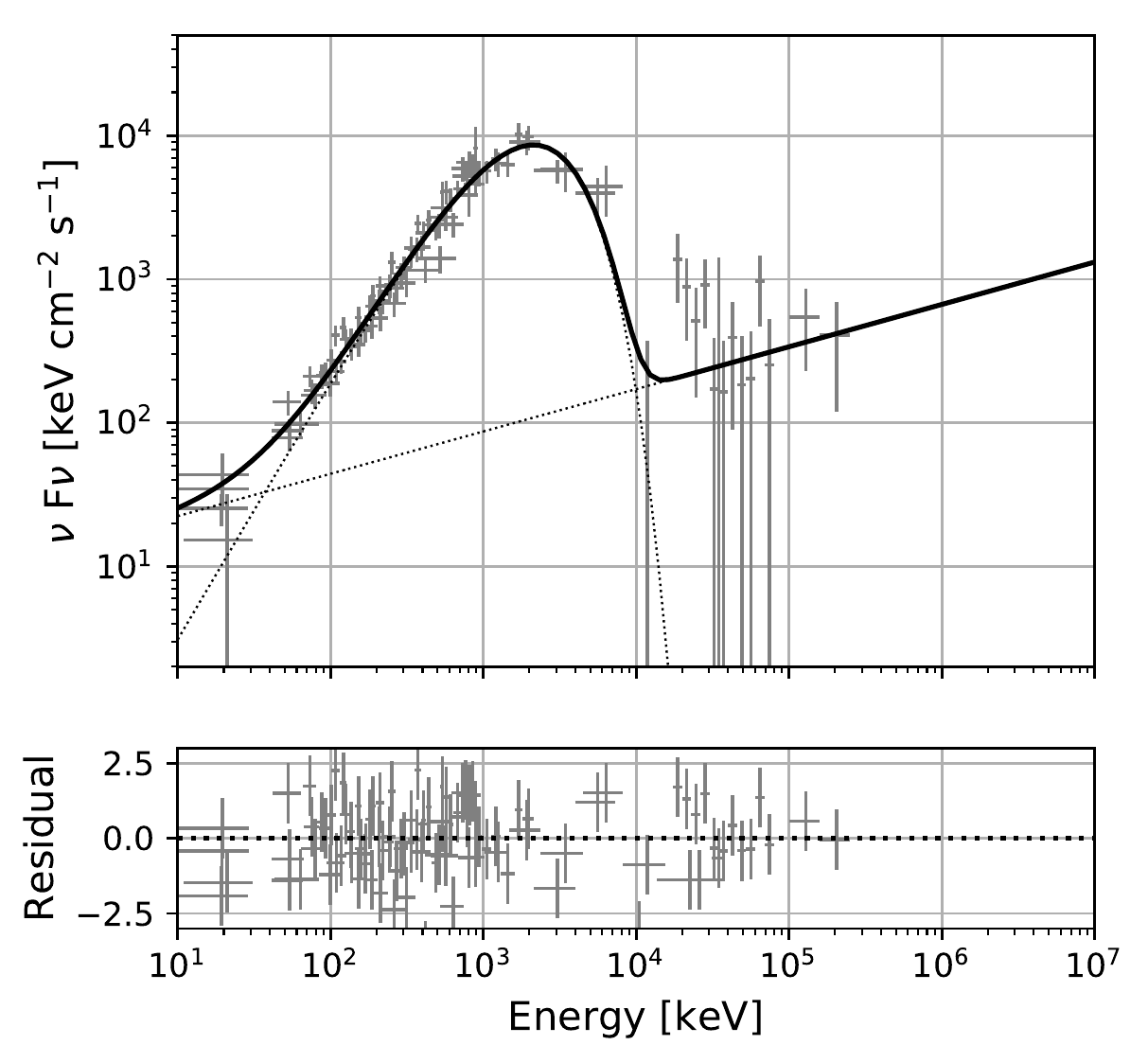}
 \includegraphics[scale=0.7]{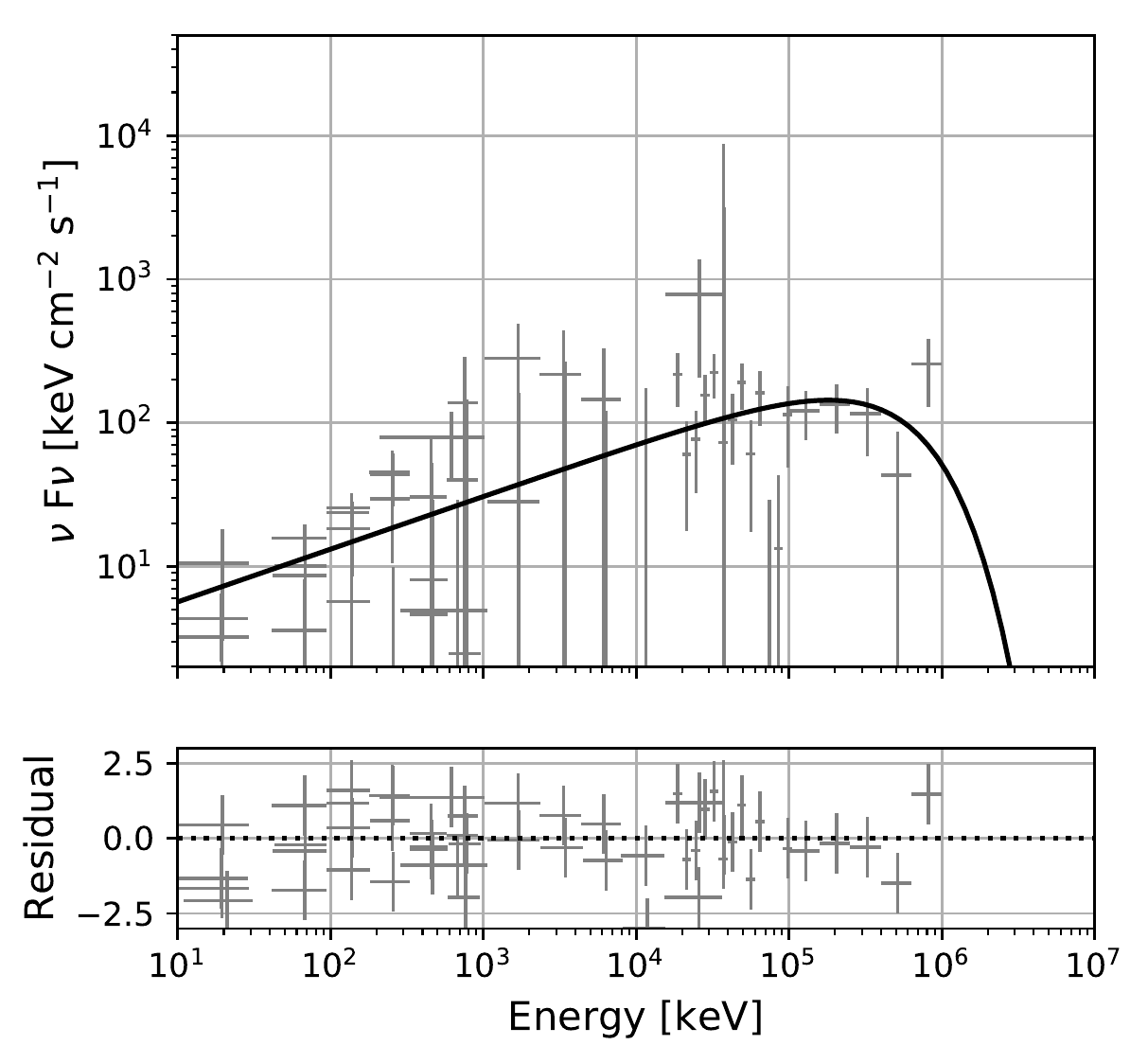}
 \caption{The spectral fit result in the three episodes: first episode (top left, [$T_{0}$ - 0.064 s, $T_{0}$ + 0.320 s], CPL), second episode (top right, [$T_{0}$ + 0.320 s, $T_{0}$ + 0.768 s], CPL + PL), and third episode (bottom left, [$T_{0}$ + 0.768 s, $T_{0}$ + 5.568 s], PL\textsubscript{break}). These spectral fits are based on the GBM + LLE + LAT data. In each plot, there are two panels. The upper panel shows the best-fit model (solid black line) and data (gray points) in $\nu$ F$_\nu$ vs. energy space. In the lower panel, the residual, (data-model)/error, is displayed.}
 \label{fig:fit1}
\end{figure}

\bibliographystyle{yahapj}
\bibliography{references}
\appendix
We perform the time-integrated spectral fit in the three episode. Figure~\ref{fig:fit1} shows the best-fit model for each episode and its residual.
\graphicspath{{./}{fit_figures/}}
\section{Time-integrated spectral fit results} \label{app:fit}
\newpage
\section{Time-resolved spectral fit results in the second episode} \label{app:fit2}
\begin{figure}[h]
 \includegraphics[scale=0.49]{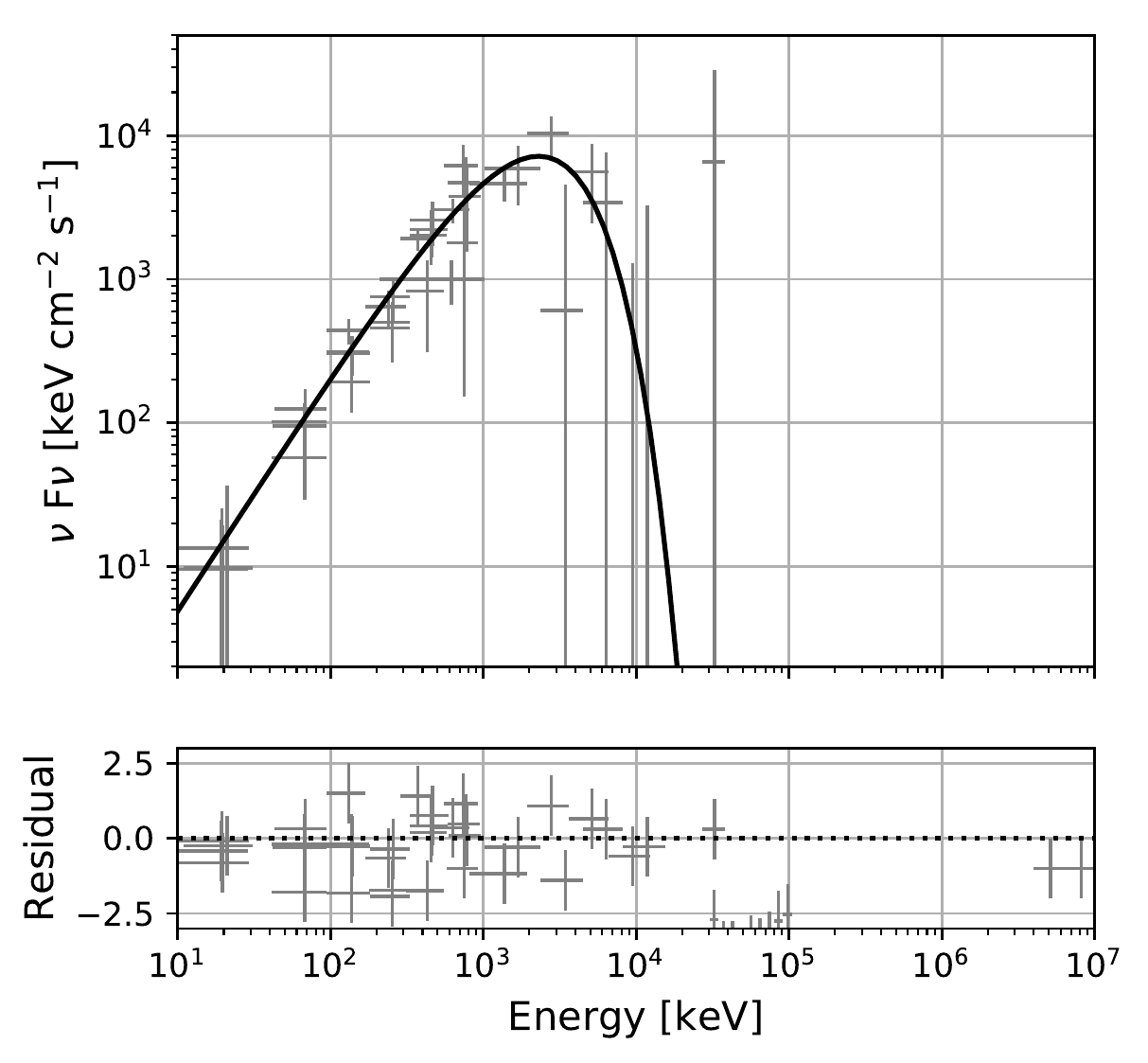}
 \includegraphics[scale=0.49]{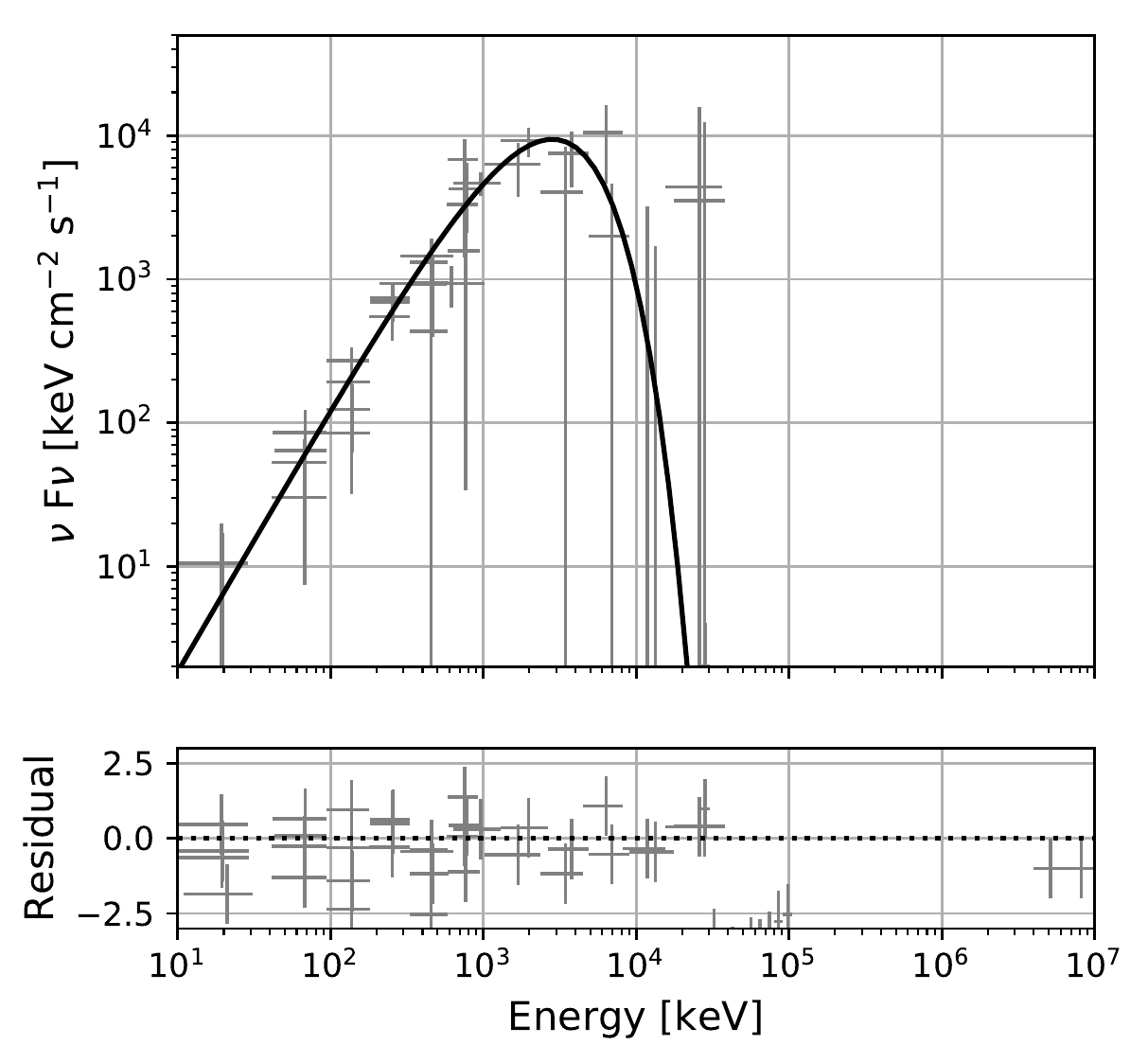}
 \includegraphics[scale=0.49]{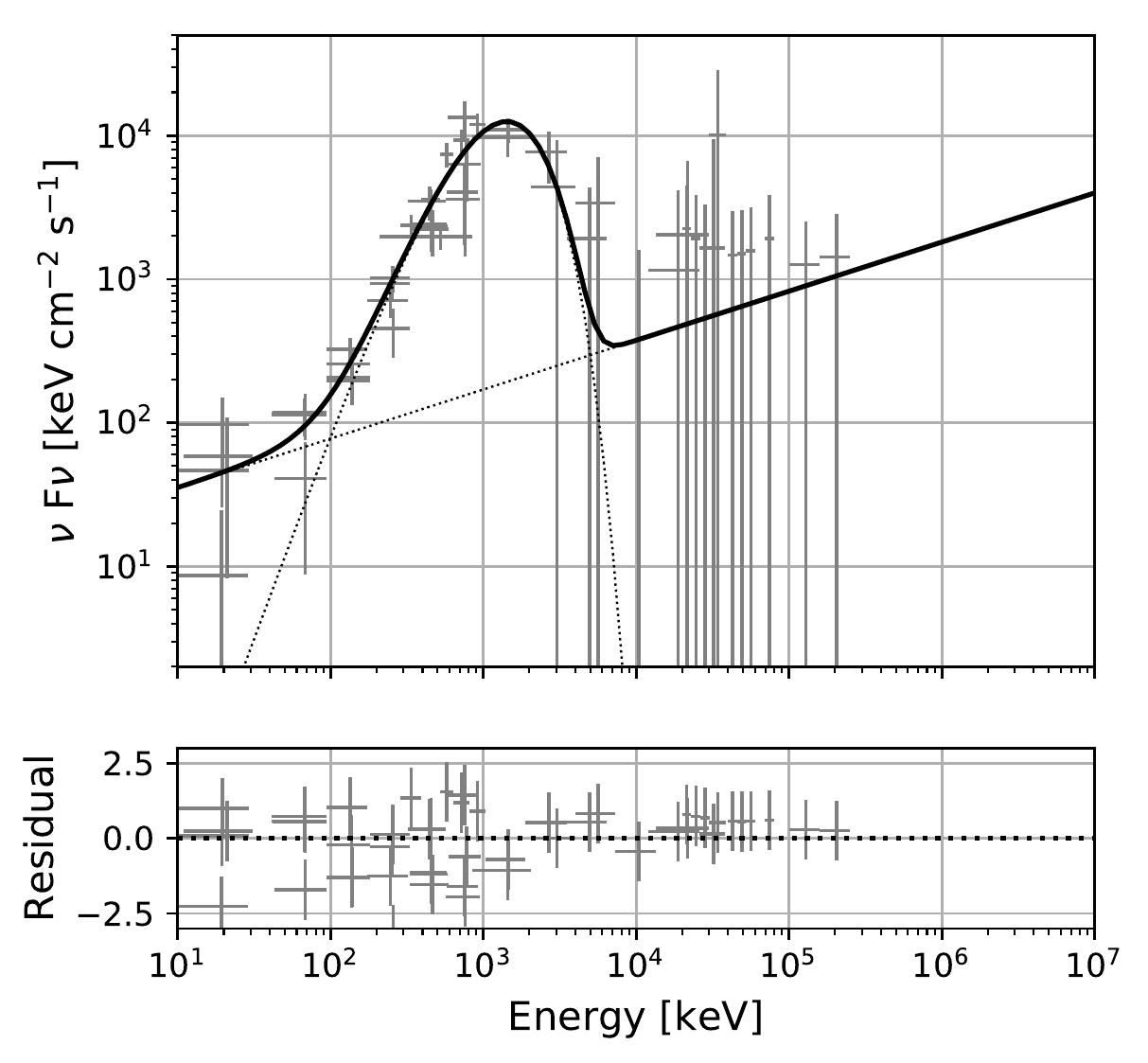}
 \includegraphics[scale=0.49]{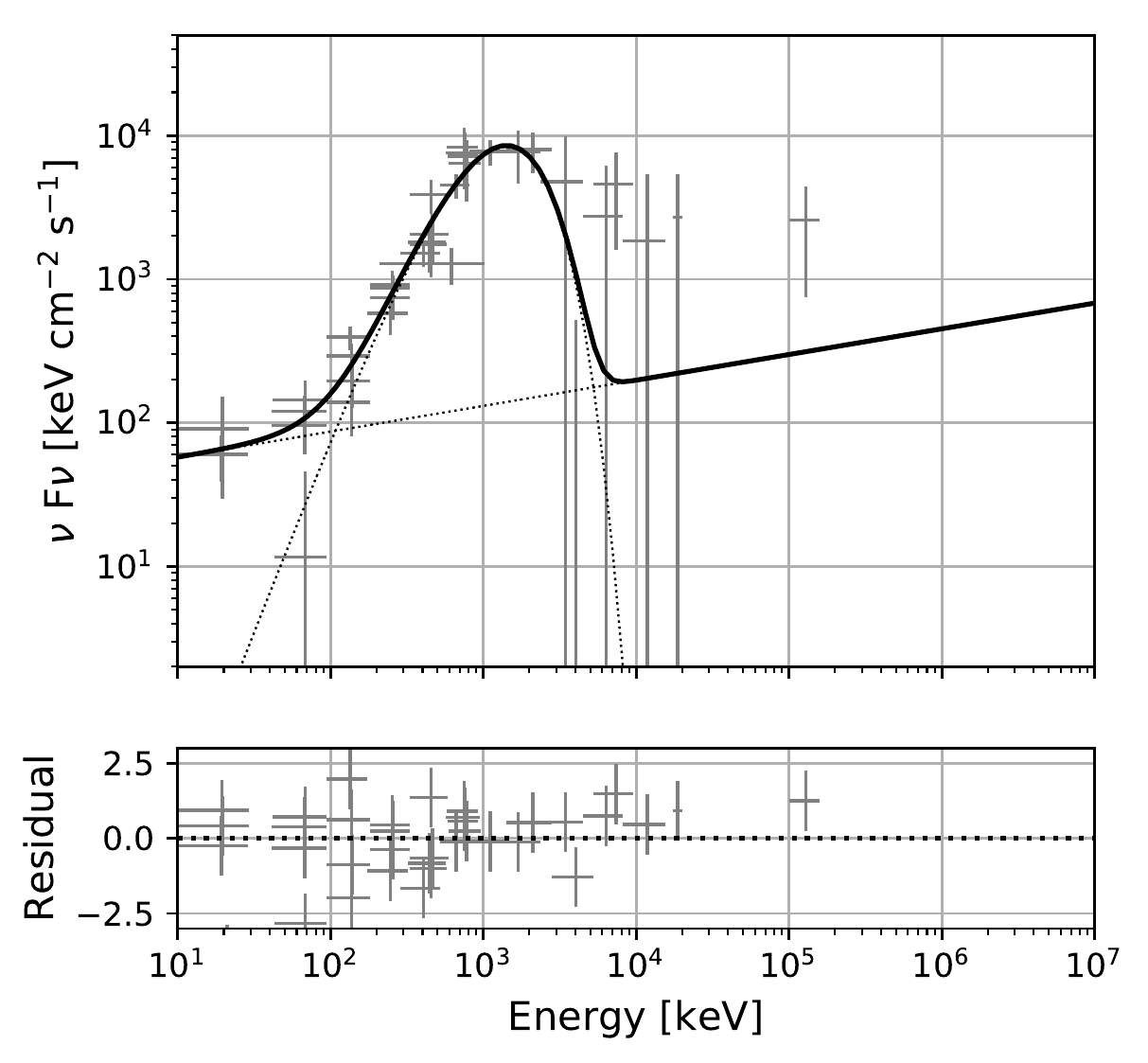}
 \includegraphics[scale=0.49]{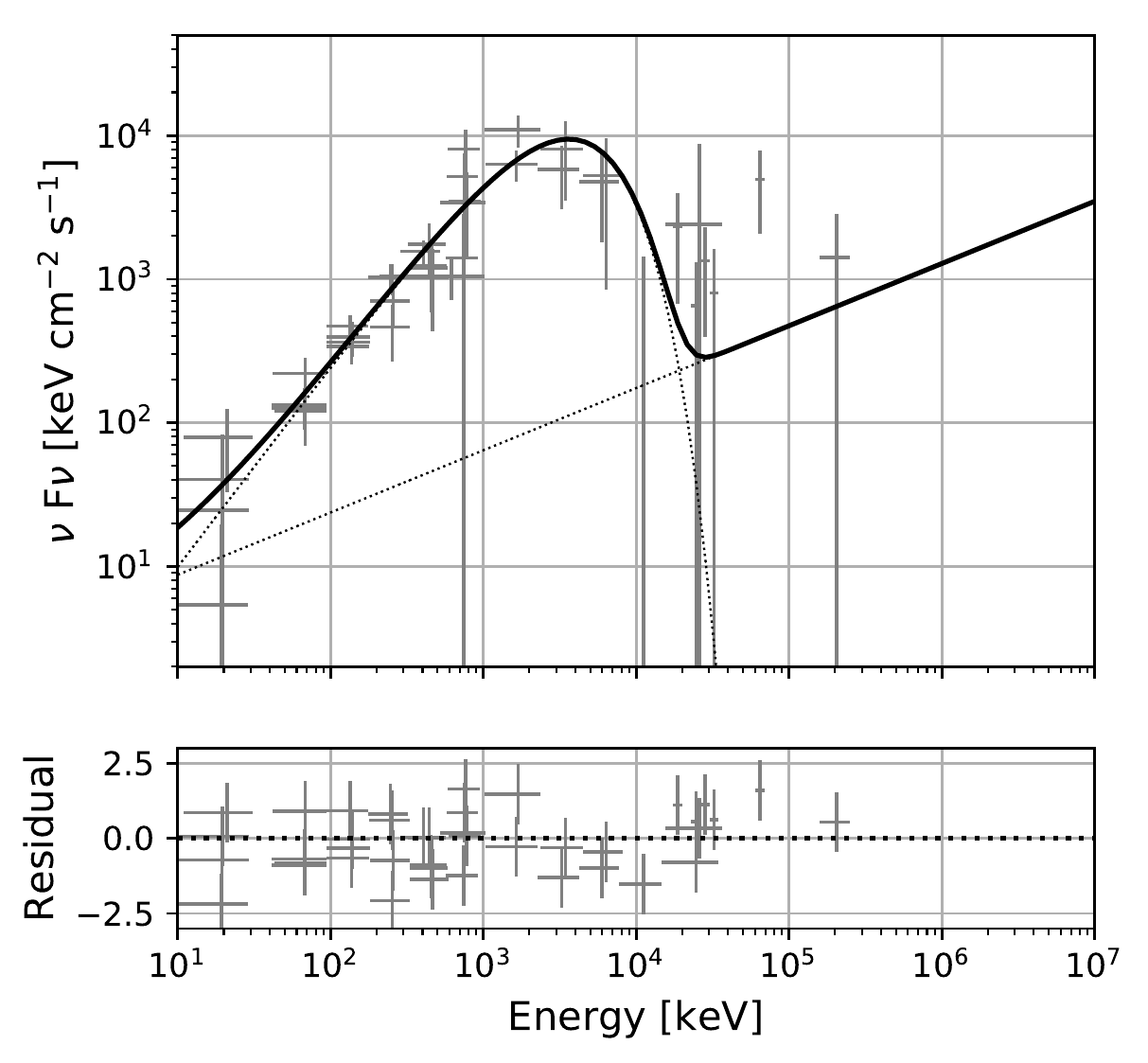}
 \includegraphics[scale=0.49]{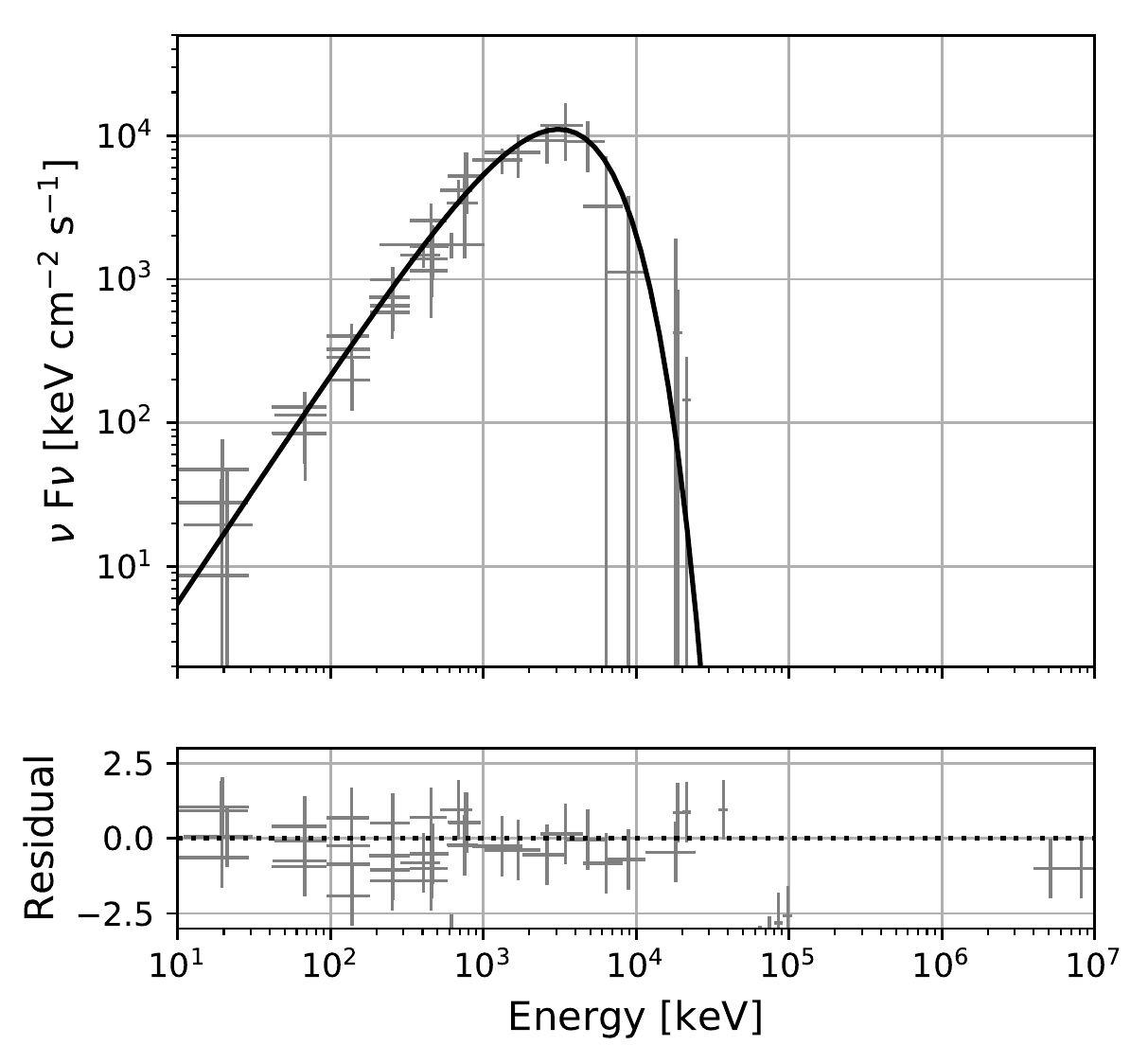}
 \includegraphics[scale=0.49]{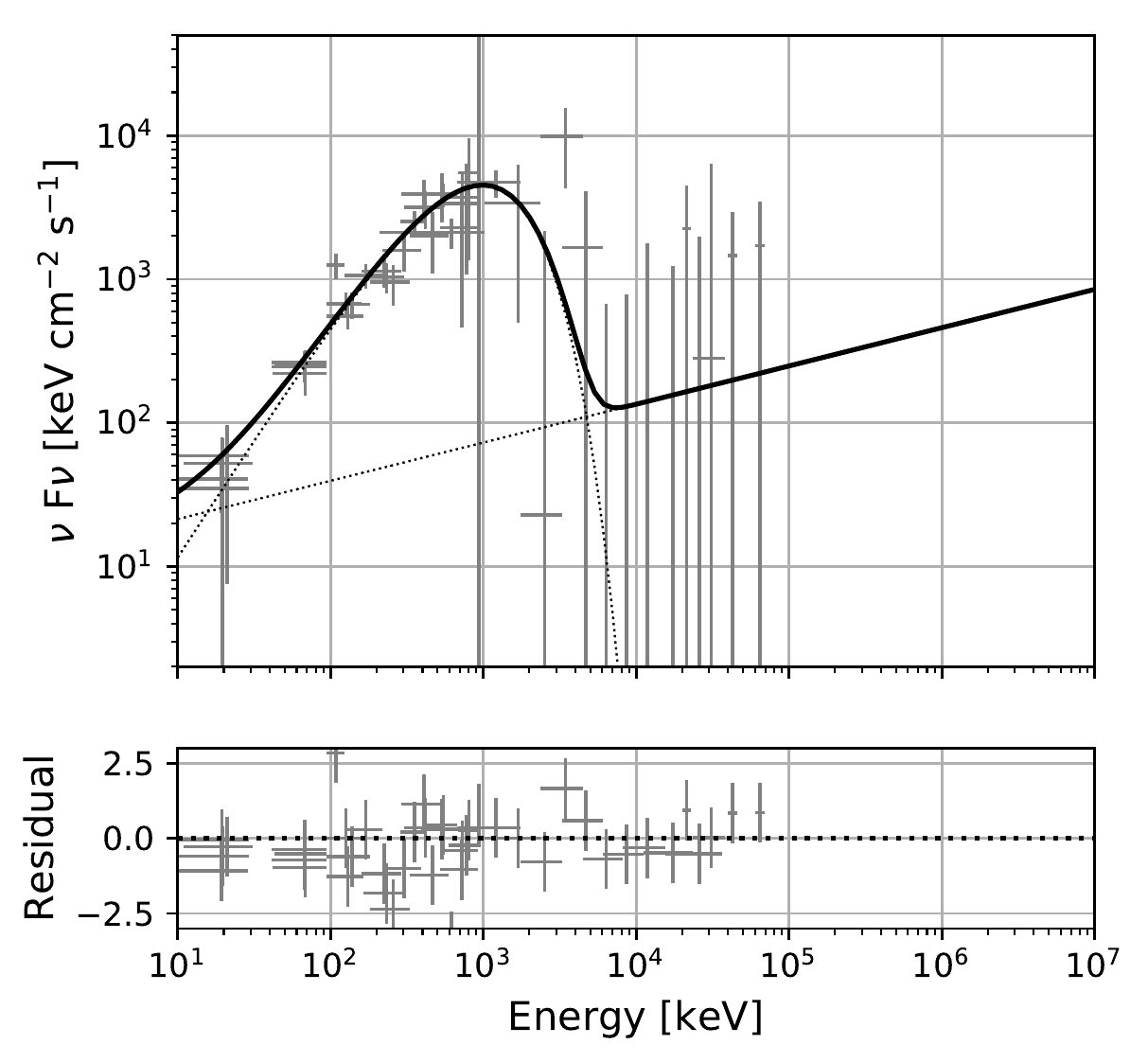}
 \caption{The 64 ms time-resolved spectral fit results of the second episode. The second episode, [$T_{0}$ + 0.320 s, $T_{0}$ + 0.768 s], is divided into seven time intervals of 64 ms time each. Their spectral fits are based on the GBM + LLE + LAT data. They are displayed in order from left to right starting from the top. In each plot, there are two panels. The upper panel shows the best-fit model (solid black line) and data (gray points) in $\nu$ F$_\nu$ vs. energy space. In the lower panel, the residual, (data-model)/error, is displayed.}
 \label{fig:fit4}
\end{figure}

We perform the time-resolved spectral analysis in the second episode. Figure~\ref{fig:fit4} shows the best-fit model for each time interval and its residual.
\end{document}